\documentclass[twocolumn]{aastex631}

\usepackage{amsmath}
\usepackage{array}
\usepackage{booktabs}
\usepackage{multirow}
\usepackage{rotating}
\usepackage{xcolor}

\begin{document}

\title{ReveaLLAGN 1: JWST Emission-Line Spectra Reveal Low-Luminosity AGN with UV-Deficient SEDs and Warm Molecular Gas}

\author[0000-0002-7743-9906]{Kameron Goold}
\affiliation{Department of Physics \& Astronomy, University of Utah, James Fletcher Building, 115 1400 E, Salt Lake City, UT 84112, USA}

\author[0000-0003-0248-5470]{Anil Seth}
\affiliation{Department of Physics \& Astronomy, University of Utah, James Fletcher Building, 115 1400 E, Salt Lake City, UT 84112, USA}

\author[0000-0001-8440-3613]{Mallory Molina}
\affiliation{Department of Physics \& Astronomy, University of Utah, James Fletcher Building, 115 1400 E, Salt Lake City, UT 84112, USA}
\affiliation{Department of Physics \& Astronomy, Vanderbilt University, Nashville, TN 37235, USA}

\author[0009-0004-9457-2495]{David Ohlson}
\affiliation{Department of Physics \& Astronomy, University of Utah, James Fletcher Building, 115 1400 E, Salt Lake City, UT 84112, USA}

\author[0000-0001-9738-3594]{Nischal Acharya}
\affiliation{Centro de Estudios de F\'{i}sica del Cosmos de Arag\'{o}n (CEFCA), Plaza San Juan 1, E–44001, Teruel, Spain}

\author[0000-0002-5666-7782]{Torsten B\"oker}
\affiliation{European Space Agency, c/o STScI, 3700 San Martin Drive, Baltimore, MD 21218, USA}

\author[0000-0003-0234-3376]{Antoine Dumont}
\affiliation{Max-Planck-Institut f{\"u}r Astronomie, K{\"o}nigstuhl 17, D-69117, Heidelberg, Germany}

\author[0000-0002-3719-940X]{Michael Eracleous}
\affiliation{Department of Astronomy \& Astrophysics and Institute for Gravitation and the Cosmos, The
Pennsylvania State University, 525 Davey Lab, University Park, PA 16802, USA}

\author[0000-0001-9490-899X]{Juan Antonio Fern\'{a}ndez-Ontiveros}
\affiliation{Istituto di Astrofisica e Planetologia Spaziali (INAF–IAPS), Via Fosso del Cavaliere 100, I–00133 Roma, Italy}
\affiliation{Centro de Estudios de F\'{i}sica del Cosmos de Arag\'{o}n (CEFCA), Plaza San Juan 1, E–44001, Teruel, Spain}

\author[0000-0001-5802-6041]{Elena Gallo}
\affiliation{Department of Astronomy, University of Michigan, 1085 S. University Ave., Ann Arbor, MI 48109, USA}

\author[0000-0003-4700-663X]{Andy D. Goulding}
\affiliation{Department of Astrophysical Sciences, Princeton University, Princeton, NJ 08544, USA}

\author[0000-0002-5612-3427]{Jenny E. Greene}
\affiliation{Department of Astrophysical Sciences, Princeton University, Princeton, NJ 08544, USA}

\author[0000-0001-6947-5846]{Luis C. Ho}
\affiliation{Kavil Institute for Astronomy and Astrophysics, Peking University, Beijing 100871, China}
\affiliation{Department of Astronomy, School of Physics, Peking University, Beijing 100871, China}

\author[0000-0002-6922-2598]{Nadine Neumayer}
\affiliation{Max-Planck-Institut f{\"u}r Astronomie, K{\"o}nigstuhl 17, D-69117, Heidelberg, Germany}

\author[0000-0002-7092-0326]{Richard M. Plotkin}
\affiliation{Department of Physics, University of Nevada, Reno, NV 89557, USA}
\affiliation{Nevada Center for Astrophysics, University of Nevada, Las Vegas, NV 89154, USA}

\author[0000-0002-3585-2639]{Almudena Prieto}
\affiliation{Universidad de La Laguna (ULL), Dpto. Astrof\'{i}sica, Avd. Astrof\'{i}sico Fco. S\'{a}nchez s/n, 38206 La Laguna, Tenerife, Spain}
\affiliation{Instituto de Astrof\'{i}sica de Canarias (IAC), C/V\'{i}a L\'{a}ctea s/n, 38205 La Laguna, Tenerife, Spain}
\affiliation{Universit{\"a}ts-Sternwarte, Fakult{\"a}t f{\"u}r Physik, Ludwig-Maximilians-Universit{\"a}t M{\"u}nchen, 81679 M{\"u}nchen, Germany}

\author[0000-0001-8557-2822]{Jessie C. Runnoe}
\affiliation{Department of Physics \& Astronomy, Vanderbilt University, Nashville, TN 37235, USA}

\author[0000-0003-2277-2354]{Shobita Satyapal}
\affiliation{George Mason University, Department of Physics and Astronomy, MS3F3, 4400 University Drive, Fairfax, VA 22030, USA}

\author[0000-0003-4546-7731]{Glenn van de Ven}
\affiliation{Department of Astrophysics, University of Vienna, T\"urkenschanzstra{\ss}e 17, 1180 Vienna, Austria}

\author[0000-0002-1881-5908]{Jonelle L. Walsh}
\affiliation{George P. and Cynthia W. Mitchell Institute for Fundamental Physics and Astronomy, Department of Physics \& Astronomy, Texas A\&M University, 4242 TAMU, College Station, TX 77843, USA}

\author[0000-0003-3564-6437]{Feng Yuan}
\affiliation{Center for Astronomy and Astrophysics and Department of Physics, Fudan University, Shanghai 200438, People's Republic of China}

\author[0000-0002-0160-7221]{Anja Feldmeier-Krause}
\affiliation{Department of Astrophysics, University of Vienna, T\"urkenschanzstra{\ss}e 17, 1180 Vienna, Austria}

\author[0000-0002-1146-0198]{Kayhan G{\"u}ltekin}
\affiliation{Department of Astronomy, University of Michigan, 1085 S. University Ave., Ann Arbor, MI 48109, USA}

\author[0000-0002-4034-0080]{Nora L{\"u}tzgendorf}
\affiliation{European Space Agency, c/o STScI, 3700 San Martin Drive, Baltimore, MD 21218, USA}

\begin{abstract}
We present near- and mid-infrared spectra of eight Low-Luminosity Active Galactic Nuclei (LLAGN), spanning nearly four orders of magnitude in black hole mass and Eddington ratio, obtained with JWST/NIRSpec and MIRI as part of the ReveaLLAGN program along with identical archival data of Cen~A. The high spatial resolution of JWST cleanly separates AGN emission from host-galaxy contamination, enabling detections of high–ionization potential lines more than an order of magnitude fainter than previously measured. Emission-line diagnostics reveal a transition at log($L_{\rm bol}/L_{\rm Edd}$) $\sim -3.5$, where the spectral energy distribution becomes increasingly deficient in ultraviolet photons. We find that rotational H$_2$ excitation temperatures are elevated ($\sim$500~K higher) compared to both higher-luminosity AGN and star-forming galaxies, while the H$_2$(0-0)S(3)/PAH$_{11.3 \mu \mathrm{m}}$ ratios are consistent with those observed in the AGN population. We discuss the possible roles of outflows, jets, and X-ray dominated regions in shaping the interstellar medium surrounding LLAGN. Silicate emission at $\sim$10~$\mu$m, localized to the nuclear region, is detected in most ReveaLLAGN targets. This dataset offers the first comprehensive JWST-based characterization of infrared emission lines in the nuclear regions of LLAGN.

\end{abstract} 

\section{Introduction} \label{sec:intro}

Low-Luminosity Active Galactic Nuclei are a class of active galaxies characterized by their relatively low accretion rates. They are the most common type of AGN in the local Universe \citep{Ho2008}, and most AGN are thought to be in a LLAGN phase at some point in their lifetimes (e.g., \citealt{Rawlings1991, Falcke2001, DiMatteo2005, Denney2014, Schawinski2015}). 

Unlike their more luminous counterparts, which exhibit prominent thermal emission from geometrically thin, optically thick accretion disks, LLAGN are thought to host radiatively inefficient accretion flows (RIAFs) in their innermost regions \citep{Narayan1995, Yuan2014, Porth2019}. This accretion mode is associated with a truncated accretion disk and reduced far-ultraviolet thermal disk emission, as evidenced by the absence of the \lq\lq big blue bump\rq\rq\ (BBB) in their spectral energy distributions (SEDs) \citep{Ho1999, Quataert1999, Maoz2007, Eracleous2010, Fern2023, Kang2024}. In addition to the lack of a strong BBB, many LLAGN also show no evidence for a broad-line region (BLR) \citep{Elitzur2014, Kang2024}, emission from a dusty torus \citep{Plotkin2012, Muller2013}, or the characteristic 1 µm inflection in their SEDs associated with the dust sublimation radius \citep{Prieto2016}. These components are, however, commonly observed in higher-accretion AGN. 

Despite their low radiative output, LLAGN are often radio-loud \citep{Ho2001, Ho2002} with compact radio cores hosting parsec scale jets \citep{Nagar2005, Mezcua2014, Baldi2018, Baldi2021}. Broadband SEDs of representative LLAGN (e.g., M87, NGC~1052, Sombrero, Cen~A, and NGC~1097), when sampled homogeneously on parsec scales, are not consistent with predictions from RIAF emission alone \citep[e.g.,][]{Meisenheimer2007, Prieto2016, Reb2018, Fernandez2012, Fern2019, Fern2023}. These works show that LLAGN SEDs can be accurately reproduced when jet emission, from the radio through the ultraviolet, is the dominant emission source, highlighting that, in the low-radiative efficiency regime, the primary channel of energy release is mechanical rather than radiative. Mechanical feedback through RIAF-launched winds \citep{Yuan2012, Yuan2015, Wang2013, Cheung2016, Park2019, Shi2021}, jet-driven outflows \citep{Falcke1995, Falcke2000, Nagar2005, Markoff2008, Maitra2011, Prieto2016}, or a combination of both, may play a large role in regulating star formation and suppressing cooling in massive galaxies \citep{Croton2006, Weinberger2017}.

LLAGN have historically been challenging to study as their faint nuclear emission is easily overwhelmed by the surrounding stellar light and dust of the host galaxy. JWST, with its 6.5-meter mirror and advanced spectroscopic instruments such as the Medium Resolution Spectroscopy (MRS) mode of the Mid-Infrared Instrument (MIRI) and the Near Infrared Spectrograph (NIRSpec), provides unprecedented sensitivity in the IR. Short exposures ($\sim$10 ks) can reach depths comparable to 2 Ms from Chandra Deep Field North (\citealt{Xue2016}, assuming the \citealt{Asmus2015} relation between mid-IR and X-ray emission). The spatial resolution of the MIRI/MRS and NIRSpec integral field unit (IFU) allows LLAGN emission to be separated from host galaxy light, and the spectral resolution enables detailed studies of line profiles that were previously not possible \citep{Goold2024}.

Infrared observations are valuable for studying LLAGN \citep{Sajina2022}, because dust that obscures the nucleus at optical and UV wavelengths emits strongly in the IR. The energy output of many AGN peaks in the mid-IR and X-ray bands \citep{Prieto2010}, with 12 $\mu$m emission tightly correlated with 2–10 keV X-ray luminosity \citep{Asmus2015}. Strong emission lines are prominent at IR wavelengths and provide valuable probes of the AGN ionizing spectrum \citep{Satyapal2008, Goulding2009}

The Revealing LLAGN (ReveaLLAGN) project uses JWST MIRI/MRS and NIRSpec IFU observations to study seven nearby LLAGN spanning a wide range of black hole masses ($10^{5.6}$–$10^{9.8}$ M$\odot$) and Eddington ratios (log($L_{\rm bol}/L_{\rm Edd}$) from $-6.2$ to $-2.7$). The first ReveaLLAGN paper, \citet{Goold2024}, presented MIRI/MRS data for the first two targets, Sombrero and NGC~1052, illustrating the power of JWST for isolating nuclear emission in LLAGN.

In this follow-up, we build on our previous analysis by presenting emission line measurements derived from nuclear spectra of all seven ReveaLLAGN targets, along with measurements from publicly available data on Centaurus~A (Cen~A). We explore unique and notable emission line signatures and investigate how these features probe the ionizing continuum and feedback processes in LLAGN. Section~\ref{sec:data} provides an overview of the data acquisition and reduction processes. Section~\ref{sec:methods} describes the spectral extraction and emission line fitting, with the resulting measurements presented in Section~\ref{sec:result}. In Section~\ref{sec:emission_line_discussion}, we explore various aspects of the emission line properties, including a discussion on probing the ionizing continuum of our targets with high-ionization potential (IP) emission lines in Section~\ref{sec:ionizing_continuum}, the impact of kinetic feedback processes on molecular gas in Section~\ref{sec:kin_mode_feedback_discussion}, and anomalous broad silicate emission in Section~\ref{sec:silicate_discussion}. Finally, our conclusions are summarized in Section~\ref{sec:conclusion}. 

\section{Data } \label{sec:data}
\begin{figure*}
  \centering
  \includegraphics[width=\textwidth]{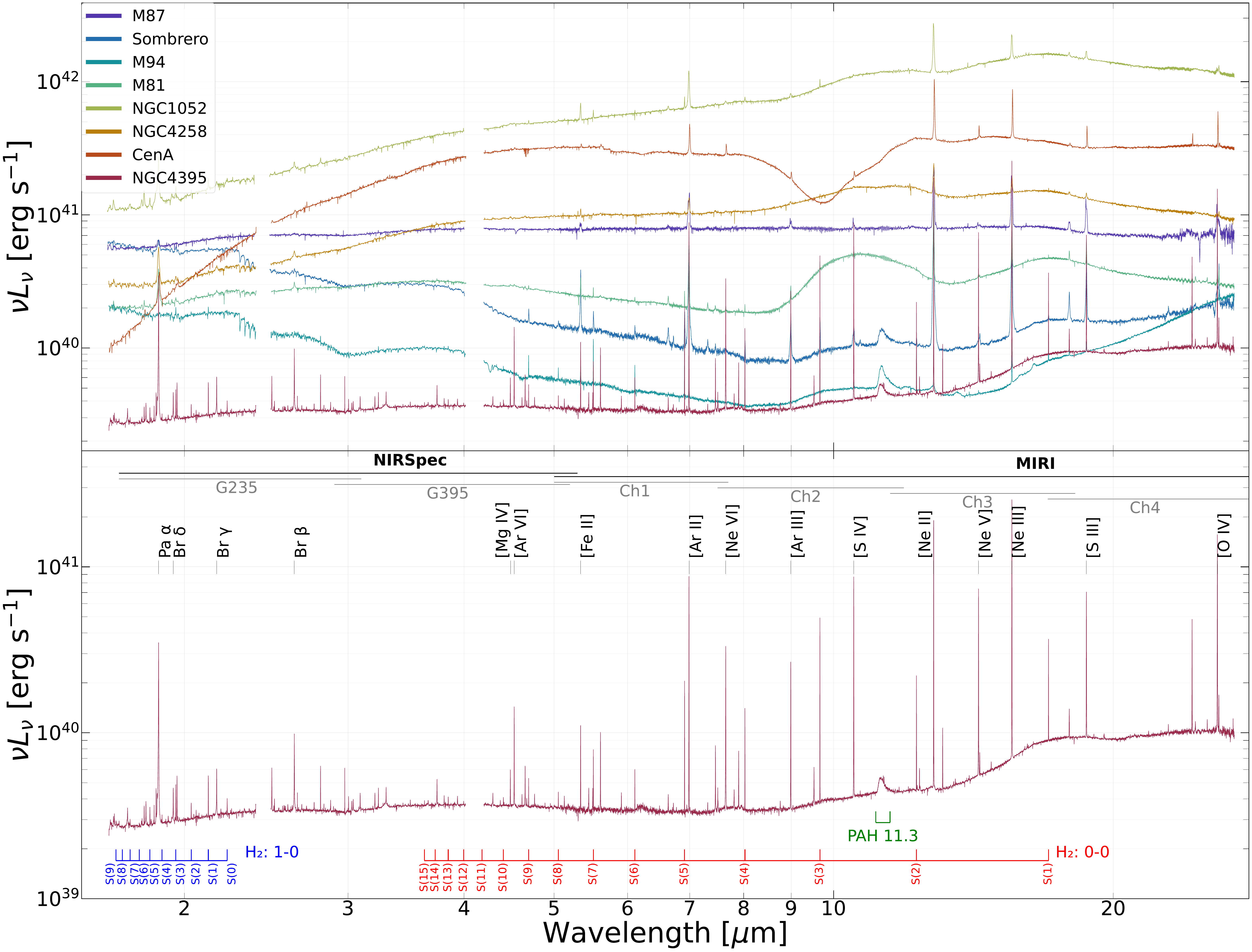}
  \caption{\textit{Top} -- Nuclear extracted spectra for ReveaLLAGN targets and Cen~A. Spectra have been normalized to MIRI/MRS channel 1. Targets are listed in the legend and colored according to Eddington Ratio (see Table~\ref{tab:galprop}). \textit{Bottom} -- NGC~4395 spectrum is shown. JWST instrument filters are labeled and prominent emission lines that are used in this work are marked. }
  \label{fig:nuc_spectra}
\end{figure*}
The ReveaLLAGN survey (Cycle 1 GO proposal, ID 2016, PI: A. Seth) observed seven nearby LLAGN, with observations occurring between 2022 and 2023. Table~\ref{tab:galprop} lists properties of the seven ReveaLLAGN survey targets as well as Cen~A, a LLAGN observed during Cycle 1 under GTO proposal ID 1269 (PI: N. Luetzgendorf). We collected data using JWST NIRSpec \citep{Jakobsen2022, Boeker2022} and MIRI/MRS \citep{Wells2015}, covering a wavelength range from 1.66 $\mu$m to 28.9 $\mu$m. 

For NIRSpec (R$\sim$2700), we used the high-resolution instrument configurations F170LP/G235H (1.66–3.17 $\mu$m) and F290LP/G395H (2.87–5.27 $\mu$m). NIRSpec has a $3\arcsec \times 3\arcsec$ field of view and a spaxel size of 0.1$\arcsec$. A physical gap between the NIRSpec detectors creates a gap in wavelength coverage, affecting the spectral ranges 2.361–2.492 $\mu$m for the G325H grating and 3.983–4.203 $\mu$m for the G395H grating. Of note; for the redshifts range of our sample Brackett-$\alpha$ (4.05 $\mu$m) falls within the G395H gap making it inaccessible in our observations. Each ReveaLLAGN target was observed for a total of 875s in each NIRSpec gratings with exposures divided into four separate dithers to improve subpixel sampling. 

As described in \citet{Argyriou23}, the MIRI/MRS IFU data span four different channels: ch1 (4.9–7.65 $\mu$m), ch2 (7.51–11.71 $\mu$m), ch3 (11.55-18.02 $\mu$m), and ch4 (17.71–28.1 $\mu$m). Each channel is further divided into sub-bands: short, medium, and long. The field of view increases for each channel: ch1 ($3.2\arcsec \times 3.7\arcsec$), ch2 ($4.0\arcsec \times 4.8\arcsec$) , ch3 ($5.2\arcsec \times 6.2\arcsec$) , and ch4 ($6.6\arcsec \times 7.7\arcsec$), and spaxel size: ch1 ($ 0.13\arcsec$), ch2 ($ 0.17\arcsec$), ch3 ($ 0.2\arcsec$), and ch4 ($ 0.35\arcsec$). The spectral resolution varies by wavelength \citep{Argyriou23},
\begin{equation}
R_{\rm{MIRI}} = 4603 - 128\frac{\lambda}{[\mu \rm{m}]}.
\end{equation}
As with the NIRSpec data we used a uniform observing strategy for MIRI/MRS in all 7 ReveaLLAGN targets: four dithers obtained in each of the three sub-bands resulting in a uniform total exposure time of 899s across the full wavelength range. Cen~A was observed in the same manner.

The JWST data presented in this paper were obtained from the Mikulski Archive for Space Telescopes (MAST) at the Space Telescope Science Institute. The complete ReveaLLAGN observations data set can be accessed via DOI: \href{https://archive.stsci.edu/doi/resolve/resolve.html?doi=10.17909/pg86-j196}{10.17909/pg86-j196}. 

\begin{deluxetable*}{l r@{\,±\,}l r@{\,±\,}l cccccccc}
\tabletypesize{\footnotesize}
\setlength{\tabcolsep}{0.03in}
\tablewidth{0pt}
\tablecaption{Galaxy Properties\label{tab:galprop}}
\tablehead{
\colhead{Galaxy Name} &
\multicolumn{2}{c}{Distance} &
\multicolumn{2}{c}{$V_{sys}$} &
\colhead{log($M_{\mathrm{Gal}})$} &
\colhead{Morph.} &
\colhead{AGN Type} &
\colhead{log($M_\mathrm{{BH}}$)} &
\colhead{$M_\mathrm{{BH}}$ Method} &
\colhead{log($L_{2-10 keV}$)} &
\colhead{$L_\mathrm{bol}$} &
\colhead{log($L_{\mathrm{bol}}/L_{\rm{edd}}$)} \\
\colhead{} &
\multicolumn{2}{c}{(Mpc)} &
\multicolumn{2}{c}{(km~s$^{-1}$)} &
\colhead{(M$_\odot$)} &
\colhead{} &
\colhead{} &
\colhead{(M$_\odot$)} &
\colhead{} &
\colhead{(erg~s$^{-1}$)} &
\colhead{(erg~s$^{-1}$)} &
\colhead{}
}
\startdata
NGC~1052  & 19.41 & 2.41 & 1560 & 3 & 10.70 & E4    & L1.9  & 8.82 & M-$\sigma_{\ast}$ & 41.52 & 42.71 & $-$4.21 \\
NGC~4258  & 7.58 & 0.08 & 461 & 0 & 10.66 & SAB(s)bc & S1.9  & 7.60 & Dyn. & 40.99 & 42.18 & $-$3.53 \\
NGC~4395  & 4.76 & 0.02 & 319 & 1 & 8.39 & SA(s)m  & S1.8  & 5.60 & Dyn. & 39.93 & 41.12 & $-$2.58 \\
M81    & 3.70 & 0.16 &$-$39 & 3 & 10.70 & SA(s)ab  & S1.5  & 7.81 & Dyn. & 40.25 & 41.44 & $-$4.47 \\
M87    & 16.70 & 0.60 & 1284 & 5 & 11.32 & cD0-1 pec & L2   & 9.81 & Dyn. & 40.66 & 41.85 & $-$6.06 \\
M94    & 4.41 & 0.08 & 308 & 1 & 10.40 & (R)SA(r)ab & L2   & 6.83 & Dyn. & 38.46 & 39.65 & $-$5.28 \\
Sombrero  & 9.55 & 0.31 & 1089 & 1 & 11.21 & SA(s)a  & L2   & 8.83 & Dyn. & 39.98 & 41.17 & $-$5.76 \\
Cen~A   & 3.68 & 0.06 & 547 & 5 & 10.67 & S0 pec  & BL Lac & 7.74 & Dyn. & 41.92 & 43.11 & $-$2.73 \\
\enddata
\tablerefs{ {\bf Distances}: NGC~1052 -- \citet{Tonry2001}; NGC~4258 -- \citet{Reid2019}; NGC~4395, M81, M94, and Cen~A -- \citet{Karachentsev2004} \tablenotemark{$\dagger$} ; M87 -- \citet{Blakeslee2009}; Sombrero -- \citet{McQuinn2016}. {\bf Systemic Velocities $V_{sys}$}: NGC~1052 -- \citet{Koss2022}; NGC~4258 -- \citet{Pesce2018}; NGC~4395 -- \citet{Reid2019,Haynes1998}; M81 -- \citet{Speights2012}; M87 -- \citet{Cappellari2011}; M94 -- \citet{Springbob2005}; Sombrero -- \citet{Sutter2022}; Cen~A -- \citet{CORV}. {\bf Galaxy Mass}: From \citet{Bi2020}. {\bf Morphological Type}: From \citet{RC3}, {\bf AGN Type}: Cen~A -- \citet{Chiaberge2001}, all others from \citet{Ho1997}, {\bf BH Mass}: NGC~1052 is based on velocity dispersion \citet{Koss2022}, NGC~4258 -- \citet{Reid2019}; NGC~4395 -- \citet{denBrok2015}; M81 and M94 -- \citet{Kormendy2013}; M87 -- \citet{EHT2019}; Sombrero -- \citep{Jardel2011}; Cen~A -- \citep{Cappellari2009}. {\bf 2-10~keV X-ray Luminosity}: 2-10~keV luminosities for Sombrero, NGC~1052, NGC~4258, NGC~4395, M81, M87, and Cen~A from \citet{Asmus2015}, M94 from \citet{Gonzalez-martin2009}, using distances in this table in all cases. {\bf Bolometric Luminosity}: All sources calculated from listed $L_{2-10kev}$ and Equations~\ref{eq:Kx_Lx},\ref{eq:lbol}. {\bf Eddington Ratio}: Calculated using listed $L_{\rm{bol}}$ and $L_{\rm{Edd}}$ from Equation~\ref{eq:ledd}. }
\tablenotetext{\dagger}{Updated values from \href{https://relay.sao.ru/lv/lvgdb/}{Catalog \& Atlas of the LV Galaxies}}
\end{deluxetable*}

\subsection{Data Reduction}\label{sec:data_reduction}

\subsubsection{NIRSpec Reduction}\label{sec:nirspec}
We processed the NIRSpec data cubes using JWST Pipeline version 1.15.0 \citep{jwstpipeline} and reference file jwst\_1256.pmap. NIRSpec IFU spectra exhibit noise, or ``wiggles'', due to spatial under-sampling at all wavelengths \citep{Law2023, Dumont2025}. These wiggles are especially prominent near compact sources like AGN. These wiggles are not fully corrected by the JWST pipeline, requiring additional post-processing for accurate spectral analysis. To address residual under-sampling effects, the NIRSpec data cubes are processed using the WIggle Corrector Kit for NIRSpec Data: WICKED. The WICKED code is a Python tool that employs Fourier analysis to identify affected spaxels and spectral template fitting to remove artifacts, preserving spectral integrity and improving measurement accuracy (see \cite{Dumont2025} for details).

\subsubsection{MIRI/MRS Reduction}\label{sec:miri}
Starting with the raw rates files, we processed MIRI/MRS data using JWST Pipeline v1.15.1 and jwst\_1293.pmap. We used the default settings, with the exception of changing the background subtraction method during calwebb\_spec2 to a pixel-by-pixel subtraction. The JWST pipeline allows multiple background files to be provided simultaneously; these are combined and then subtracted from the science data. For NGC~4258, NGC~4395, M81, and M87, we supplied all four background observations for each target, allowing the pipeline to combine them, which improved the signal-to-noise of the corresponding nuclear spectra by 2–15\%. We used individual background observations for Sombrero, NGC~1052, and M94.

\subsection{Nuclear Spectra Extraction}\label{sec:nuclear_extraction}

The spatial resolution of JWST allows us to cleanly extract high-quality LLAGN nuclear spectra from their host galaxies \citep{Goold2024}. However, the point spread function (PSF) of both NIRSpec and MIRI/MRS varies with wavelength. To accurately define an optimal aperture that captures the PSF at each wavelength, we empirically measured encircled energy profiles as a function of radius using the spectrophotometric standard stars; 2MASS J17571324+6703409 (Program ID: 3399, PI: M. Perrin) and HD192163 (Program ID: 1031, PI: A. Labiano) for NIRSpec and MIRI/MRS, respectively. Encircled energy profiles are measured out to 1.2\arcsec, and the radius enclosing 75\% of the total flux at each wavelength is adopted as the nuclear aperture radius for spectral extraction. Nuclear spectra are extracted from each science target using this wavelength dependent aperture. This corresponds to an aperture of 0.13\arcsec\ radius at 1.7~$\mu$m and 0.61\arcsec\ radius at 26~$\mu$m. In physical units, these apertures correspond to nuclear extraction regions ranging from approximately 2–11 pc for the nearest galaxies in the sample (M81 and Cen~A) to 12–58 pc for the most distant (NGC~1052). We performed sky subtraction using an annulus with an inner and outer radius corresponding to 3–5 times the 75\% encircled energy radius. We create a 1D spectrum by averaging all spatial pixels in the annulus, scale it by the aperture area, and then subtract it from the summed spectrum extracted from the aperture. Sky subtraction has little effect on hydrogen recombination or forbidden emission lines but can affect extended emission, particularly molecular hydrogen and PAH features. Effects on molecular hydrogen are discussed further in Section~\ref{sec:measure_h2_excitation}. Figure~\ref{fig:nuc_spectra} shows the extracted nuclear spectra for each ReveaLLAGN target, normalized to the MIRI/MRS Ch1 flux levels. This normalization (performed to provide consistent flux scaling across instruments and channels) results in a median flux variation of 3\% across all MIRI/MRS channels and 10\% across NIRSpec filters. The largest changes are seen in Sombrero (37\% decrease in NIRSpec filter G235, and 19.4\% increase in MIRI/MRS channel 4) and M94 (14\% decrease in NIRSpec filter G235, and 24.7\% increase in MIRI/MRS channel 4). PAH measurements (Section~\ref{sec:measure_nuclear_pah}) are taken from non–sky-subtracted spectra normalized to MIRI/MRS channel 1, whereas all other emission-line measurements use sky-subtracted, unnormalized nuclear spectra. 

All reduced datacubes and extracted nuclear spectra described in this section are publicly available at the ReveaLLAGN data archive.\footnote{\url{https://reveallagn.github.io/}}

\subsection{Bolometric Corrections}\label{sec:bolometric corrections}

The bolometric luminosity (\(L_{\mathrm{bol}}\)) represents the total radiative output of an AGN integrated over all wavelengths. It can be determined either by directly integrating the full spectral energy distribution (SED) or by applying wavelength-specific bolometric corrections to monochromatic or band-limited luminosities, such as those measured in the X-ray or optical bands. X-ray bolometric corrections are known to depend on both luminosity and Eddington ratio \citep{Marconi2004, Vasudevan2007, Eracleous2010, Nemmen2014, Duras2020, Gupta2024}. 

For several nearby low-luminosity AGN, including NGC~1052, M87, Cen~A, and the Sombrero galaxy, bolometric luminosities derived from high-resolution SED integrations are available in \citet{Prieto2010, Prieto2014, Fern2023}. However, to ensure consistency across our entire sample and to facilitate direct comparisons to supplemental data, we adopt a uniform bolometric correction based on the luminosity-dependent 2--10~keV X-ray relation from \citet{Duras2020}. This correction, derived from a sample of roughly 1000 Type~1 and Type~2 AGN with multiwavelength coverage, provides a well-calibrated and widely applicable proxy for \(L_{\mathrm{bol}}\):

\begin{equation}
K_{\rm{X}}(L_{\rm{2-10keV}}) = a \left[ 1 + \left( \frac{\log \left( \frac{L_{\rm{2-10keV}}}{L_\odot} \right)}{b} \right)^c \right],
\label{eq:Kx_Lx}
\end{equation}

where \(a = 15.33\), \(b = 11.48\), and \(c = 16.2\).

When compared with the SED-integrated bolometric luminosities of NGC~1052, M87, Cen~A, and the Sombrero galaxy, the luminosities derived using the \citet{Duras2020} relation show a median absolute difference of 0.54~dex. Given this offest and for methodological consistency, we adopt the X-ray–based bolometric correction throughout this work to enable a uniform and interpretable comparison across the full sample.

\subsection{Supplemental and Comparison Data}\label{sec:supplemental_data}

In this paper, we analyze emission lines and line ratios, and investigate their correlations with galaxy properties like X-ray luminosity and Eddington ratio, defined as \(L_{\mathrm{bol}} / L_{\mathrm{Edd}}\), where 
\begin{equation}\label{eq:ledd}
L_{\mathrm{Edd}} = 1.26 \times 10^{38} \left( \frac{M_{\mathrm{BH}}}{M_\odot} \right)~\mathrm{erg\,s^{-1}}, 
\end{equation}
and:
\begin{equation}\label{eq:lbol}
  L_{\rm{Bol}} = K_{\rm{X}}(L_{\rm{2-10keV}}) \times L_{\rm{2-10keV}}.
\end{equation}

For the ReveaLLAGN sample, we adopt absorption-corrected 2--10~keV X-ray luminosities from \citet{Asmus2015}, except for M94, where we use the value reported by \citet{Gonzalez-martin2009} (see Table~\ref{tab:galprop}). X-ray luminosities can vary significantly depending on modeling assumptions, particularly in the treatment of intrinsic absorption (i.e., the adopted column density). The \citet{Asmus2015} catalog was selected for its rigorous quality control: sources with low counts, inconsistent measurements, or indications of Compton-thick obscuration are excluded, and multiple robust measurements from different instruments are averaged when available. Distances, and black hole masses are compiled from various literature sources, as detailed in Table~\ref{tab:galprop}.  X-ray luminosities are scaled to these distances, and Eddington ratios are computed using Equations~\ref{eq:Kx_Lx}, \ref{eq:ledd} and \ref{eq:lbol}.

To place the ReveaLLAGN emission-line measurements in a broader context, we compare them with a compilation of archival spectroscopic AGN samples that include key mid-infrared (MIR) fine-structure lines. These surveys provide well-calibrated fluxes for transitions such as [OIV]$_{26 \mu \rm{m}}$, [NeV]$_{14 \mu \rm{m}}$, [NeIII]$_{15 \mu \rm{m}}$, and [NeII]$_{12 \mu \rm{m}}$ \citep{Sturm2002, Goulding2009, Dudik2009, Tommasin2010, Fernandez2016}. For comparisons based purely on MIR emission lines, we assemble a sample of 186 archival AGN drawn from these works.

We examine the relationship between [OIV]$_{26 \mu \rm{m}}$ and [NeV]$_{14 \mu \rm{m}}$ luminosities, adopting distances from \citet{Fernandez2021}. After cross-matching the individual survey catalogs, we end up with 112 AGN.

We further explore the relation between [NeV]$_{14 \mu \rm{m}}$ luminosity and the ratios [NeV]$_{14 \mu \rm{m}}$/[NeII]$_{12.6 \mu \rm{m}}$ and [NeV]$_{14 \mu \rm{m}}$/[NeIII]$_{15.6 \mu \rm{m}}$ against both 2–10 keV X-ray luminosities and Eddington ratios. Black hole masses for the comparison sample are taken from \citet{Fernandez2021}, while absorption-corrected X-ray luminosities are from \citet{Asmus2015} and scaled using distances from \citet{Fernandez2021} to stay consistent between all plots, as discussed in Sect.~\ref{sec:bolometric corrections}. All X-ray luminosities are converted to bolometric values using the same correction factor applied to the ReveaLLAGN. Eddington ratios are calculated in the same manner as the ReveaLLAGN sample using Equations~\ref{eq:Kx_Lx}, \ref{eq:ledd} and \ref{eq:lbol}.

After combining these datasets, the resulting comparison sample contains 43 AGN with consistent MIR, X-ray, and black hole mass measurements. These supplemental datasets were selected to ensure well-documented measurement methods and calibration standards and maximize the size of the comparison sample.

\section{Emission Feature Measurements}\label{sec:methods}

The high signal-to-noise nuclear spectra of the ReveaLLAGN sources contain numerous emission lines with complex profiles. We construct line lists using the emission-line-rich nuclear spectrum of NGC~4395 as a reference. NGC~4395 is a dwarf galaxy and hosts the lowest-mass black hole in our sample, displaying strong emission lines over a relatively low continuum. Emission features are identified through visual inspection of the nuclear spectrum of NGC~4395 followed by the other targets. Lines are verified using the NIST Atomic Spectra Database, as well as line lists provided with \texttt{CLOUDY} \citep{Ferland2017}. The final line list includes 131 emission lines spanning NIRSpec and MIRI, and include hydrogen recombination lines, molecular hydrogen transitions, and atomic fine-structure lines. The bottom half of Figure~\ref{fig:nuc_spectra} shows the extracted nuclear spectrum of NGC~4395 and a selection of prominent emission lines used in this work.

We fit emission lines using a multi-step procedure, following the same method described in \citet{Goold2024}. First, we mask each emission line using a 2000 km/s window centered on its rest-frame wavelength. To determine the local continuum, we require at least 100 unmasked spectral elements on either side of the line center. A linear function is fit to this segment and subtracted from the unmasked spectral data. We define the fitting window by locating the points on either side of the rest wavelength where the flux falls to 5\% of its peak value at line center. If the window spans fewer than 30 spectral elements on either side, we extend it accordingly and adjust the boundaries to maintain symmetry about the rest wavelength. We fit both single- and multi-Gaussian models to the continuum-subtracted data: the single-Gaussian fit is used to assess whether a line is detected and resolved, while multi-Gaussian models are used to measure line fluxes and velocity structure.

We estimate uncertainties using a Monte Carlo approach. For each emission line, we calculate the standard deviation of the flux within the continuum window. This value is used as the representative noise level and serves as the standard deviation for a normal distribution from which we draw the random perturbations. The fitting procedure is repeated multiple times on these noise-perturbed spectra to derive uncertainties. The wavelength solution, FLT-8, associated with our pipeline version has a 1$\sigma$ wavelength calibration error of ~10 km~s$^{-1}$. We adopt this as the minimum uncertainty for velocity-based measurements. We consider an emission line detected if the integrated flux of the best-fit single-Gaussian exceeds $3\sigma$; otherwise, we report the 3$\sigma$ upper limit (Appendix~\ref{sec:appendix_table}). We classify lines as resolved if their full width at half maximum (FWHM) exceeds the width of the instrumental line spread function (LSF). The instrumental FWHM is given by
\begin{equation}
\rm{FWHM}_{\rm{LSF}} = c/R,
\end{equation}
where c is the speed of light in km/s and R is the spectral resolution of the instrument (Section~\ref{sec:data}). 
The measured FWHM values are corrected for instrumental broadening using
\begin{equation}
\rm{FWHM}_{\rm{Corrected}} = \sqrt{\rm{FWHM}_{\rm{Measured}}^2 - \rm{FWHM}_{\rm{LSF}}^2}.
\end{equation}

The multi-Gaussian model is a flexible parameterization that can include up to ten components. For each emission line, we fit models containing between one and ten Gaussians. For a given number of components ($N_{\mathrm{Gauss}}$), we randomly perturb the initial parameters and refit the line ten times, retaining the fit with the lowest Bayesian Information Criterion (BIC) for that $N_{\mathrm{Gauss}}$. We then compare these best-fit models across all values of $N_{\mathrm{Gauss}}$ and adopt the model with the overall lowest BIC as the final fit. 

We use the multi-Gaussian models solely to describe the complexity of high--signal-to-noise line profiles; individual Gaussians are not assumed to represent distinct physical components of the AGN. All derived quantities---line fluxes, peak velocities, and FWHM---are measured from the final best-fit multi-Gaussian model (see Appendix~\ref{sec:appendix_table}).

In many cases, broad or nearby emission lines are blended with one or more neighboring lines. We de-blend these using constrained multi-Gaussian models. Blended emission lines are identified when the rest wavelength of one line falls within 2$\sigma$ of the central wavelength of a neighboring line’s best-fit single-Gaussian model. Within each blended group, we select a primary line based on scientific interest, and classify the remaining lines as secondary. We constrain the parameters of secondary lines using unblended reference lines with high S/N ratios and similar ionization potentials. For example; the group of blended lines at 14.32 $\mu$m consists of a primary line, [NeV], and a secondary line, [ClII]. The shape (but not amplitude) of the secondary line is constrained based on a reference line, [FeII] 5.34 $\mu$m, while the primary line is freely fit with up to ten Gaussian components \citep[see Fig.~2 of][for an example of this process]{Goold2024}.

\subsection{Excitation Temperatures of Molecular Hydrogen \texorpdfstring{(H$_2$)}{(H2}}\label{sec:measure_h2_excitation}

\setlength{\tabcolsep}{3pt}
\begin{deluxetable}{l r@{\,±\,}l r@{\,±\,}l r@{\,±\,}l r@{\,±\,}l r@{\,±\,}l r@{\,±\,}l}
 \tabletypesize{\scriptsize}
 \tablecaption{H$_2$ Excitation Temperatures in ReveaLLAGN Targets\label{tab:excite_temp}}
 \tablewidth{0pt}
 \tablehead{
 \colhead{\textbf{Galaxy}} &
 \multicolumn{2}{c}{\textbf{T$_{3,1}$}} &
 \multicolumn{2}{c}{\textbf{T$_{4,2}$}} &
 \multicolumn{2}{c}{\textbf{T$_{5,3}$}} &
 \multicolumn{2}{c}{\textbf{T$_{6,4}$}} &
 \multicolumn{2}{c}{\textbf{T$_{7,5}$}} &
 \multicolumn{2}{c}{\textbf{T$_{8,6}$}} \\
 \colhead{} &
 \multicolumn{2}{c}{(K)} &
 \multicolumn{2}{c}{(K)} &
 \multicolumn{2}{c}{(K)} &
 \multicolumn{2}{c}{(K)} &
 \multicolumn{2}{c}{(K)} &
 \multicolumn{2}{c}{(K)}
 }
 \startdata
 NGC~1052  & \multicolumn{2}{c}{--} & 800 & 30 & 880 & 20 & \multicolumn{2}{c}{--} & 1240 & 20 & \multicolumn{2}{c}{--} \\
 NGC~4258  & 360 & 10 & \multicolumn{2}{c}{--} & 770 & 30 & \multicolumn{2}{c}{--} & 1330 & 600 & \multicolumn{2}{c}{--} \\
 NGC~4395  & 400 & 10 & 500 & 10 & 620 & 10 & 690 & 10 & 870 & 50 & \multicolumn{2}{c}{--} \\
 M81    & 560 & 40 & 670 & 30 & 880 & 50 & 1050 & 30 & 1270 & 20 & 1880 & 110 \\
 M87    & \multicolumn{2}{c}{--} & \multicolumn{2}{c}{--} & \multicolumn{2}{c}{--} & \multicolumn{2}{c}{--} & \multicolumn{2}{c}{--} & \multicolumn{2}{c}{--} \\
 M94    & 390 & 10 & 570 & 10 & 820 & 10 & 870 & 10 & 1120 & 20 & 1480 & 150 \\
 Sombrero  & 540 & 10 & 770 & 20 & 870 & 10 & 1070 & 40 & 1330 & 30 & \multicolumn{2}{c}{--} \\
 Cen~A   & \multicolumn{2}{c}{--} & \multicolumn{2}{c}{--} & \multicolumn{2}{c}{--} & \multicolumn{2}{c}{--} & \multicolumn{2}{c}{--} & \multicolumn{2}{c}{--} \\
 \enddata
\end{deluxetable}

Following the approach outlined by \citet{Lambrides2019}, we derive molecular hydrogen (H$_2$) excitation temperatures from pure rotational transitions in the mid-infrared. We use line pairs of the same parity and assume that the level populations are governed by a Boltzmann distribution under conditions of local thermal equilibrium. The upper-level column density for each transition is determined via:

\begin{equation} N_{J+2} = \frac{4 \pi D^{2}_{L} F_{J}}{A_{J+2 \rightarrow J} (E_{J+2}-E_{J})} \end{equation}

Here, $F_J$ is the observed line flux, $D_L$ is the luminosity distance, $A_{J+2 \rightarrow J}$ is the Einstein A-coefficient for the respective transition, and E$_{J+2}-E_{J}$ is the difference in energy levels. Einstein coefficients and energy levels are taken from \cite{Turner1977} and \cite{Roueff2019}. Since the extraction aperture is wavelength dependent and the H$_2$ emission is spatially extended, we normalized all H$_2$ line fluxes to a common aperture size equal to that used for the 0--0~S(1) transition ($\sim0.6^{\prime\prime}$). This was done by scaling each measured flux by the ratio of its value in the native aperture to that measured in a larger 1.2$^{\prime\prime}$ aperture, under the assumption that the spatial distribution of the emission is roughly constant between these aperture sizes. 
We calculate the excitation temperature between the upper ($u$) and lower ($l$) levels (T$_{u,l}$) from the Boltzmann equation:

\begin{equation} T_{u,l} = \frac{E_u - E_l}{\ln \left( \frac{N_l}{N_u} \cdot \frac{g_u}{g_l} \right)}, \end{equation}
where \( g \) is the statistical weight of the energy level, accounting for its degeneracy. For molecular hydrogen, \( g_J = (2J + 1) \) for para-H$_2$ (even \( J \)) and \( g_J = 3(2J + 1) \) for ortho-H$_2$ (odd \( J \)). Excitation temperatures are provided in Table~\ref{tab:excite_temp} and discussed in Section~\ref{sec:kin_mode_feedback_discussion}. 

As noted in Section~\ref{sec:data_reduction}, sky subtraction affects measurements of extended emission such as PAH and molecular hydrogen. Although sky subtraction reduces the flux of H$_2$ emission, the resulting differences in H$_2$ excitation temperatures between sky-subtracted and non-sky-subtracted spectra are smaller than 1$\sigma$ of the uncertainties reported in Table~\ref{tab:excite_temp}.

\subsection{PAH Flux and Equivalent Width}\label{sec:measure_nuclear_pah}

Polycyclic aromatic hydrocarbons (PAHs) are complex molecules that absorb UV/optical photons and re-emit in the mid-infrared \citep{Draine2007A}. Their main emission bands arise from vibrational modes, including C–C stretching (6.2, 7.7 $\mu$m), C-H in-plane bending (8.6 $\mu$m), and C-H out-of-plane bending (11.3, 12.7 $\mu$m) \citep{Leger1984, Allamandola1985}. The relative band strengths depend on ionization, size, and molecular structure: ionized PAHs dominate the 6.2, 7.7, and 8.6 $\mu$m features, while neutral PAHs contribute strongly at 3.3 and 11.3 $\mu$m \citep{Draine2001, Li2004, Xie2022}. Larger PAHs, with higher heat capacities, radiate at longer wavelengths than smaller PAHs \citep{Li2012}. Around AGN, PAHs are vulnerable to destruction by hard radiation fields \citep{Voit1992, Smith2007, Xie2021}, while shocks in LLAGN efficiently erode small PAH molecules \citep{ Diamond-Stanic2010, Zhang2022}. These dependencies make PAHs useful tracers of local radiation fields, shocks, and star formation activity \citep{Peeters2004, Tielens2008, Xie2018}.

We focus exclusively on the 11.3~$\mu$m feature, as it is the most isolated and robust PAH band in the mid-infrared spectra of our LLAGN. To measure the 11.3 $\mu$m PAH feature, we estimate the local continuum using a power-law function and fit the PAH emission profiles using a blended Drude model consisting of two drude components (Appendix~\ref{sec:appendix_PAH}). We calculate the integrated flux and equivalent width (EW) using:

\begin{equation}
  EW = \int_{\lambda_1}^{\lambda_2} \left( 1 - \frac{f(\lambda)}{f_c(\lambda)} \right) d\lambda,
\end{equation}

where $f(\lambda)$ is the flux of the PAH emission and $f_c(\lambda)$ is the flux of the continuum at wavelength $\lambda$. We estimate errors via a Monte Carlo fitting process with upper limits set to 3~$\sigma$. Flux and EW for the 11.3 $\mu$m PAH features are shown in Table~\ref{tab:pah_sil_params}.  

\setlength{\tabcolsep}{2pt}  
\begin{deluxetable}{l c c c c}
\tabletypesize{\scriptsize}
\tablecaption{Select PAH and Silicate Emission in ReveaLLAGN\label{tab:pah_sil_params}}
\tablewidth{0pt}
\tablehead{
\colhead{\textbf{Galaxy}} &
\colhead{\textbf{F$_{11.3}$}} &
\colhead{\textbf{EW$_{11.3}$}} &
\colhead{\textbf{H$_2$ S(3)/PAH$_{11.3\,\mu\mathrm{m}}$}} &
\colhead{\textbf{S$_{Sil}$}} \\[2pt]
\colhead{} &
\colhead{($10^{-14}$ erg s$^{-1}$ cm$^{-2}$)} &
\colhead{($\mu$m)} &
\colhead{} &
\colhead{}
}
\startdata
NGC~1052   & \multicolumn{1}{c}{0.568 ± 0.176} & \multicolumn{1}{c}{–0.007 ± 0.001} & 0.174 & –0.19  \\
NGC~4258   & \multicolumn{1}{c}{2.211 ± 0.466} & \multicolumn{1}{c}{–0.011 ± 0.002} & 0.024 & –0.282 \\
NGC~4395   & \multicolumn{1}{c}{0.791 ± 0.016} & \multicolumn{1}{c}{–0.057 ± 0.001} & 0.642 & –0.026 \\
M81        & \multicolumn{1}{c}{5.345 ± 1.033} & \multicolumn{1}{c}{–0.021 ± 0.004} & 0.044 & –0.75 \\
M87        & \multicolumn{1}{c}{$<0.302$}      & \multicolumn{1}{c}{$<0.014$}        & --    & 0.012 \\
M94        & \multicolumn{1}{c}{6.563 ± 0.050} & \multicolumn{1}{c}{-0.267 ± 0.003}  & 0.039 & –0.209 \\
Sombrero   & \multicolumn{1}{c}{0.809 ± 0.019} & \multicolumn{1}{c}{–0.181 ± 0.002}  & 0.132 & –0.182 \\
Cen~A      & \multicolumn{1}{c}{49.137 ± 2.196}& \multicolumn{1}{c}{-0.031 ± 0.001}  & --    & 0.857 \\
\enddata
\tablecomments{
$F_{11.3}$ is the integrated flux of the 11.3~$\mu$m PAH feature. 
$S_{sil}$ represents the silicate strength at 10.5~$\mu$m, except for Cen~A which is measured at 9.7~$\mu$m. 
Upper limits correspond to $3\sigma$ non-detections. 
}

\end{deluxetable}
\subsection{Broad Silicate Emission}\label{sec:measure_nuclear_silicate}

The broad silicate feature at 10~$\mu$m is seen in emission in all targets except M87 and Cen~A. M87 shows no signs of silicate features while in Cen~A the silicate feature is seen in absorption. To characterize the strength of these features, we adopt the definition from \cite{Spoon2022}:

\begin{equation}
S_{sil} = -ln \left( \frac{f(\lambda_{peak})}{f_c(\lambda_{peak})} \right),
\end{equation}
where $f(\lambda_{peak})$ is the flux density of the spectra at the peak wavelength, and $f_c(\lambda_{peak})$ is the flux density of the continuum at the peak wavelength and we use spectra normalized to MIRI/MRS channel 1 (Section~\ref{sec:data_reduction}) in units of erg/s/cm$^{2}$/$\mu$m. We estimate the local continuum using a power-law fit to the regions 8–8.5 $\mu$m and 13.5–14 $\mu$m on either side of the profile. We modify the continuum window for Sombrero, NGC~1052, and NGC~4395 where the blue edge of the window is set to a wavelength range of 8.7-8.9 $\mu$m instead. 

The peak wavelength for silicate emission in this region typically occurs at 10.5 $\mu$m \citep{sturm2005, Mason2015}. However, this region overlaps with the [S~IV] emission line in many targets. To minimize contamination, we mask a velocity window of $\pm$1000 km/s (2000 km/s for the Sombrero galaxy and M87) centered on the rest-frame wavelength of [S~IV]. This silicate feature is seen in absorption in Cen~A so we use $\lambda_{peak}$ of 9.7~$\mu$m. We then estimate $f(\lambda_{peak})$ and $f_c(\lambda_{peak})$ by averaging the adjacent 25 unmasked spectral elements (50 for the Sombrero galaxy and M87) around $\lambda_{peak}$. The strength of this silicate feature is included in Table~\ref{tab:pah_sil_params}.

\section{Nuclear Emission Results}\label{sec:result}

\subsection{Emission Line Measurements}\label{sec:nuclear_ionic}

Table~\ref{tab:number_of_emission_lines} presents the total number of unblended detected emission lines for each ReveaLLAGN target. It breaks down the counts by category, including hydrogen recombination lines, molecular hydrogen lines, and ionic lines. It also includes the median number of Gaussian components used to fit each category. The table also lists the median number of Gaussian components used to fit each category, which, while not physically meaningful, reflects the complexity of the line velocity profiles in each galaxy. With a total of 106 of these types of lines detected at S/N $>$ 3, NGC~4395 exhibits the richest emission line spectrum, while M87 displays the fewest, with only 13. Detailed measurements of these emission lines are provided in Appendix~\ref{sec:appendix_table} (Table~\ref{tab:line_results}).

\subsection{Detailed Line Profiles of Forbidden Emission Lines}\label{sec:line_profile}

\begin{figure*}
  \centering
  \includegraphics[width=\textwidth]{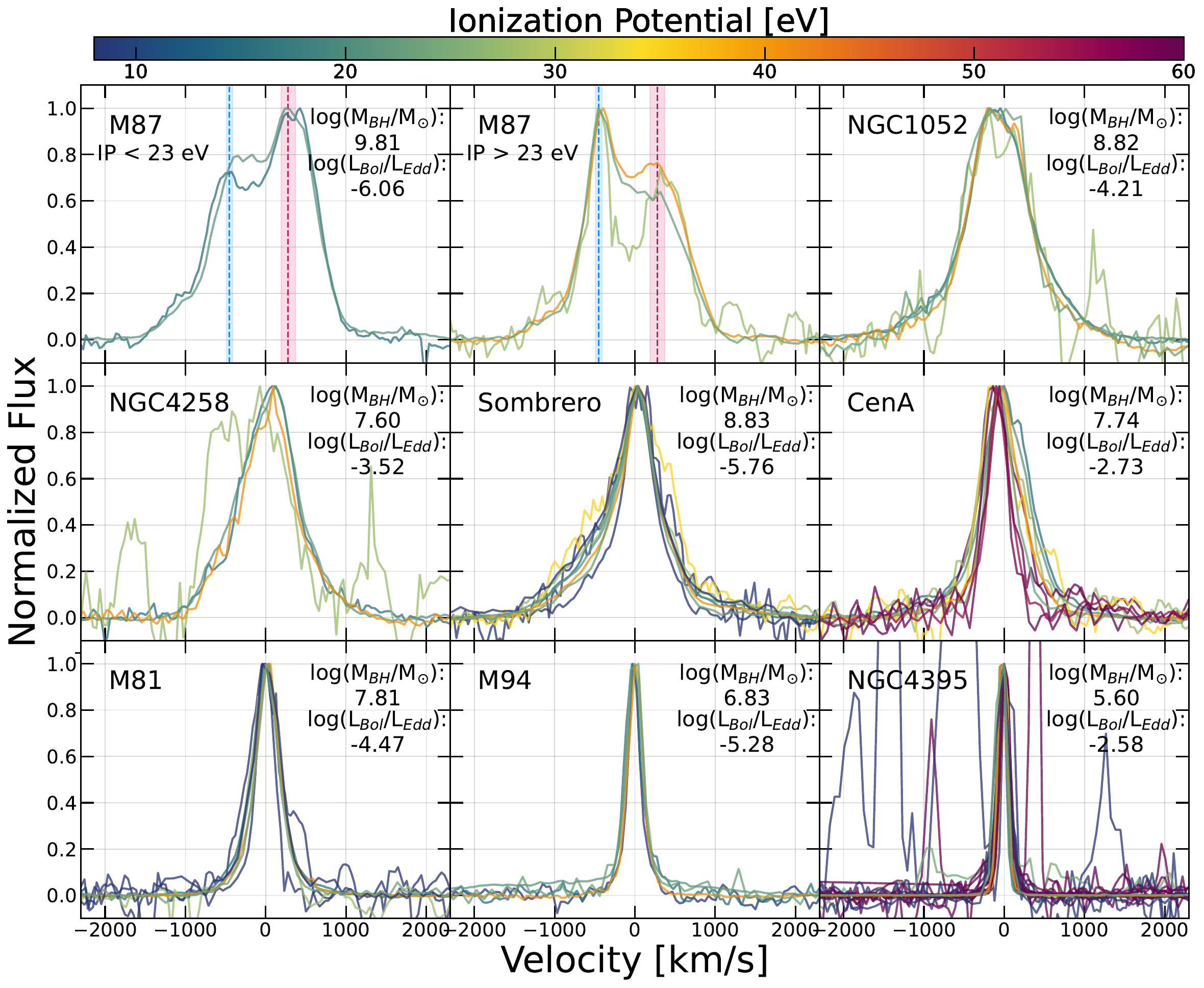}
  \caption{Forbidden emission line profiles with S/N$>$25 from each galaxy. Emission lines are centered at rest velocity and normalized to peak flux. The log Eddington ratio and log black hole mass (in solar masses) are shown in each panel. The colorbar indicates the IP of each line. The top row displays M87, where all high S/N lines are double-peaked. In the second and third columns of this row, the dominant peak shifts at an IP of 23 eV. The red (280 km/s) and blue (-450 km/s) dashed lines represent the average red and blue peaks for all high S/N sources and their 1$\sigma$ spread. The middle row shows galaxies with the next broadest lines, and the bottom row those with the narrowest.}
  \label{fig:lineprofiles}
\end{figure*}

We present normalized line profiles in Figure~\ref{fig:lineprofiles} for high signal-to-noise forbidden emission lines (S/N ratio $>$ 25). Within each galaxy, the line profiles display similar shapes, indicating they share a common origin. The variations between targets however, are quite pronounced. NGC~1052, NGC~4258, and Sombrero all exhibit broad and sometimes asymmetric line profiles (median FWHM$_{Corrected}$ of 860, 780, and 620 km/s respectively), with line widths varying systematically with IP \citep{Goold2024}. Cen~A, M81, M94, and NGC~4395 present narrower and more symmetric line profiles (median FWHM$_{Corrected}$ of 490, 390, 210, and 90 km/s respectively), while NGC~4395 displays the narrowest lines in our sample. This is perhaps not surprising, as these three galaxies are the nearest in the sample and host relatively low-mass black holes, making it unlikely that the extraction aperture encompasses a significant amount of fast-moving gas. 

A more complete interpretation of the line maps will be given by (Dumont et al., {\em in prep}), though we note here an interesting trend with ionization potential seen in M87. Every high S/N emission line in M87's nuclear spectra has a pronounced double-peaked profile. One peak is red-shifted by 280 $\pm$ 90 km/s while the other is blue-shifted by 450 $\pm$ 40 km/s, with a noticeable saddle at near the systemic velocity. Emission lines in M87 with an IP of less than 23 eV have a dominant red peak, while emission lines with an IP greater than 23 eV have a dominant blue peak. While only lines with S/N \>25 are shown in Figure~\ref{fig:lineprofiles}, other detected lines, such as [SIV]$_{10.5 \mu \rm{m}}$, [NeV]$_{14.3 \mu \rm{m}}$, and [OIV]$_{26 \mu \rm{m}}$, also display the same double-peaked structure, with red and blue peaks aligned to those shown.

A disk of ionized gas in M87 is well documented from HST observations \citep{Harms1994, Ford1994, Macchetto1997, Walsh2013}, which would be consistent with the double-peaked profiles seen in our nuclear spectra if the disk is unresolved. \cite{Osorno2023} find similar double-peaked profiles in the spectral profile of [OI] (eV $<$ 23). They find this [OI] emission to be consistent with an ionized disk at a position angle of 25$^\circ$ with overlapping contributions from biconical outflows. Our integral field results are fully consistent with this, with small (0.2\arcsec) positional shifts along a position angle of $\sim$212$^{\circ}$ between the two peaks, with the red peak being to the North~East. 

\begin{deluxetable*}{lcccccccccc}
\tablecaption{Number of Detected Emission Lines and Median Number of Gaussian Components \label{tab:number_of_emission_lines}}
\tablewidth{0pt}
\tablehead{
\colhead{Galaxy} & 
\colhead{N$_H$} & \colhead{$<$N$_{\rm Gauss,H}$$>$} &
\colhead{N$_{\mathrm{H}_{2}}$} & \colhead{$<$N$_{\rm Gauss,H_2}$$>$} &
\colhead{N$_{\mathrm{Low-IP}}$} & \colhead{$<$N$_{\rm Gauss,Low-IP}$$>$} &
\colhead{N$_{\mathrm{High-IP}}$} & \colhead{$<$N$_{\rm Gauss,High-IP}$$>$} &
\colhead{N$_{\mathrm{Total}}$} \\
\colhead{} &
\colhead{} & \colhead{} &
\colhead{} & \colhead{} &
\colhead{($<$45~eV)} & \colhead{} &
\colhead{($\geq$45~eV)} & \colhead{} &
\colhead{}
}
\startdata
NGC~1052  & 5  & 5 & 15 & 3 & 11 & 4 & 8  & 1 & 39 \\
NGC~4258  & 6  & 4 & 5  & 2 & 10 & 3 & 4  & 1 & 25 \\
NGC~4395  & 25 & 2 & 31 & 1 & 25 & 1 & 25 & 2 & 106 \\
M81       & 9  & 2 & 18 & 1 & 17 & 2 & 11 & 1 & 55 \\
M87       & 1  & 3 & 2  & 1 & 8  & 4 & 2  & 2 & 13 \\
M94       & 0  & -- & 17 & 1 & 13 & 2 & 3  & 1 & 33 \\
Sombrero  & 2  & 2 & 14 & 2 & 14 & 4 & 4  & 2 & 34 \\
Cen~A     & 3  & 3 & 1  & 1 & 11 & 2 & 13 & 1 & 28 \\
\enddata
\tablecomments{
 Columns: 
(1) Galaxy name; 
(2) Number of detected hydrogen recombination lines (N$_H$); 
(3) Median number of Gaussian components fitted to the H lines ($<$N$_{\rm Gauss,H}$$>$); 
(4) Number of detected molecular hydrogen emission lines (N$_{\mathrm{H}_{2}}$); 
(5) Median number of Gaussian components fitted to the H$_2$ lines ($<$N$_{\rm Gauss,H_2}$$>$); 
(6) Number of detected low-ionization emission lines (N$_{\mathrm{Low-IP}}$, E$_{\rm ion} <$ 45~eV); 
(7) Median number of Gaussian components for low-ionization lines ($<$N$_{\rm Gauss,Low-IP}$$>$); 
(8) Number of detected high-ionization emission lines (N$_{\mathrm{High-IP}}$, E$_{\rm ion} \geq$ 45~eV); 
(9) Median number of Gaussian components for high-ionization lines ($<$N$_{\rm Gauss,High-IP}$$>$); 
(10) Total number of detected emission lines in the spectrum (N$_{\mathrm{Total}}$).
}
\end{deluxetable*}

 \subsection{Spatial Concentration of Emission Lines}

\begin{figure}
  \centering
  \includegraphics[width=\columnwidth]{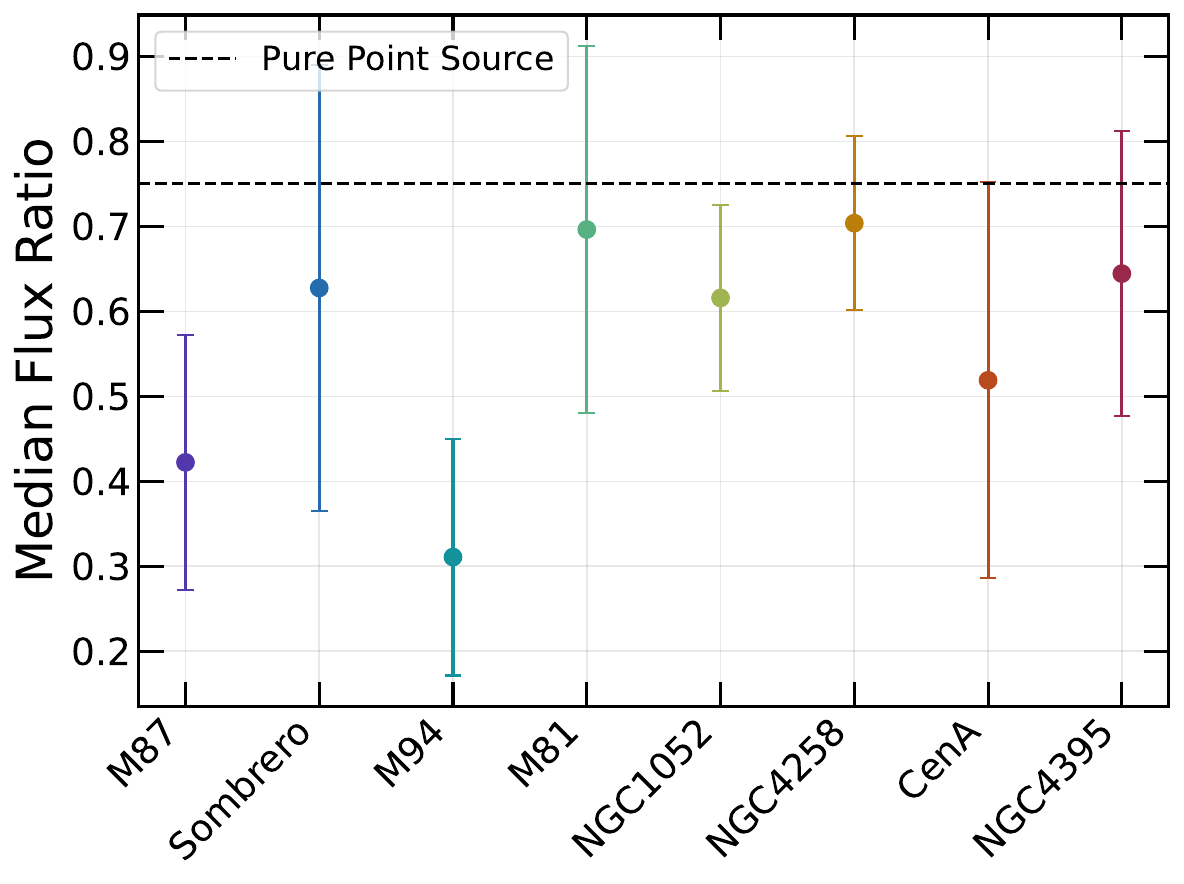}
  \caption{Emission from the nuclear extracted spectra is consistent with that of a point-like source. The dashed line marks the ratio expected for an unresolved point source. For each galaxy, we show the median and standard deviation of the flux ratios computed across all detected ionic lines. Data points are ordered, left to right, by Eddington ratio and colored using a sequential color scheme carried through the rest of the paper.}
  
  \label{fig:nuc_large_ratio}
\end{figure}

To assess whether the emission lines in the nuclear aperture spectra are consistent with an unresolved (point-like) spatial profile, we compare the fraction of flux captured by the nuclear aperture to the expected encircled-energy fraction for a point source (MIRI/MRS: HD 192163; NIRSpec: 2MASS J17571324+6703409). For each galaxy, we measure emission-line fluxes within both the nuclear aperture, defined to contain 75\% of the encircled energy of the continuum, and a larger 1.2\arcsec aperture (see Section~\ref{sec:nuclear_extraction}). We then compute the nuclear-to-total flux ratio for all unblended lines with signal-to-noise ratios greater than 3, and report the median ratio per galaxy in Figure \ref{fig:nuc_large_ratio}. Error bars indicate the standard deviation of all line measurements within each galaxy. A ratio close to 0.75 suggests that the line emission shares the same compact spatial distribution as the continuum point source.

A black dashed line denotes the 75\% encircled energy radius of a point source. With the exception of M94, which displays clear evidence of extended emission \citep{Constantin2012}, most galaxies exhibit forbidden-line emission that remains unresolved even at the spatial resolution of JWST. This suggests that the line emission is dominated by an unresolved nuclear component smaller than the extraction aperture, rather than from larger-scale host galaxy structures. Physical aperture sizes for each emission line are included in Table~\ref{tab:line_results}. The sensitivity and resolution of JWST allow us to isolate these nuclear lines and measure their fluxes accurately, minimizing contamination from surrounding host emission. Further discussion of isolating the AGN component in Sombrero (the faintest ReveaLLAGN galaxy in the IR) is presented in \citet{Goold2024}.

\section{Emission Line Analysis and Discussion}\label{sec:emission_line_discussion}

Infrared (IR) emission features offer a powerful diagnostic tool for probing the physical conditions and processes in AGN \citep{Spinoglio1992, Sajina2022}. A key advantage of IR diagnostics is their reduced sensitivity to dust extinction compared to optical and UV lines, making them especially valuable for probing the obscured regions commonly found around AGN \citep{Feltre2023}. Transitions from varying ionized states of elements such as neon, sulfur, argon, and oxygen can trace the ionizing continuum that is otherwise not observable in the far-UV and X-ray due to Galactic or intrinsic absorption \citep{Fernandez2016, Fern2023,Riffel2013}.

The infrared wavelength range includes emission from both low- and high-ionization species. High-ionization lines such as [NeV]$_{14 \mu \rm{m}}$ and [OIV]$_{26 \mu \rm{m}}$ require a hard ionizing spectrum typically associated with AGN \citep{Fern2023}, while lower-ionization lines like [NeII]$_{12 \mu \rm{m}}$ and [NeIII]$_{15 \mu \rm{m}}$ can be excited by both AGN and star-forming regions. Ratios between low- and high-ionization lines, when interpreted with photo-ionization models, are effective in disentangling the contributions from AGN activity, star formation, and shocks to the observed emission \citep{Ho2007, Zhuang2019, Feltre2023}.

Molecular hydrogen and PAH features are also prominent in the IR and serve as key tracers of the warm molecular phase and star formation activity. H$_{2}$ rotational lines probe shock-heated gas, often tracing interactions with jets or outflows \citep{Lambrides2019, Kristensen2023, Riffel2025}. While PAH emission is commonly used to measure star formation \citep{Schreiber2004, Peeters2004, Xie2019}, it is suppressed in AGN-dominated environments \citep{Zhang2022}. EW of the 6.2 $\mu$m PAH feature and ratios between H$_{2}$ and the 11.3 $\mu$m PAH feature provide useful diagnostics to determine AGN activity.

Broad silicate emission and absorption features in the mid-IR have been attributed either to the obscuring torus or to dust associated with host-galaxy structures \citep{Goulding2012}. Silicate emission, though less common, arises when the dust is optically thin, allowing direct observation of reprocessed radiation from the nuclear region.

In the following discussion, we take advantage of the available IR features (forbidden atomic lines, molecular lines, and silicate emission) to investigate the central engine, its interaction with the surrounding medium, and the spectral characteristics that distinguish LLAGN from their more luminous counterparts.

\subsection{Probing the Ionizing Continuum of LLAGN with Lines from High-IP Species}\label{sec:ionizing_continuum}

\begin{figure}
  \centering
  \includegraphics[width=\linewidth]{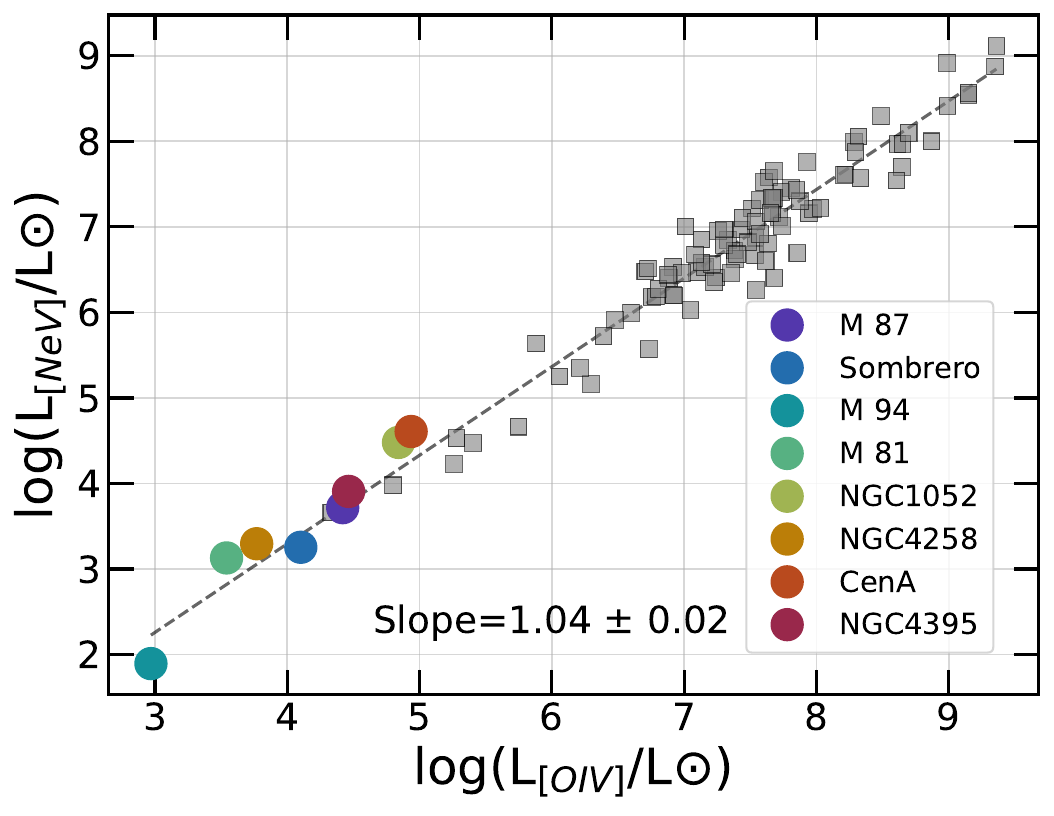}
  \caption{[NeV]$_{14 \mu \rm{m}}$ luminosity versus [OIV]$_{26 \mu \rm{m}}$. ReveaLLAGN targets all show low-luminosity detections in both lines, with M94 being the faintest. For comparison, data from previous surveys \citet{Sturm2002, Goulding2009, Tommasin2010, Fernandez2016} are shown in gray. 1$\sigma$ errors are smaller than data points. [NeV]$_{14 \mu \rm{m}}$ and [OIV]$_{26 \mu \rm{m}}$ are strongly correlated across many orders of magnitude in solar luminosity, and the ReveaLLAGN sample extends this trend to the lowest luminosities. ReveaLLAGN sources are ordered in the legend according to Eddington ratio and colored using the same sequential color map as in Figure~\ref{fig:nuc_large_ratio}.}
  \label{fig:nev_oiv}
\end{figure}

In \citet{Goold2024}, we presented the first detections of [NeV]$_{14 \mu \rm{m}}$ emission in Sombrero and NGC~1052, improving upon previous upper limits. In this work, we expand the analysis to the full ReveaLLAGN survey, detecting [NeV]$_{14 \mu \rm{m}}$ and [OIV]$_{26 \mu \rm{m}}$ in every source, along with many additional high- and low-ionization potential lines. [NeV]$_{14 \mu \rm{m}}$ is a robust diagnostic of AGN activity in the mid-infrared, as it is less affected by dust extinction compared to optical and UV diagnostics and is not subject to depletion onto grains \citep{McKaig2024}.

Since [NeV] requires photons with energies above 97.1~eV, it cannot be produced in significant amounts by star formation \citep{Abel2008}. 
In standard, high-accretion AGN, this flux is largely produced by the inner accretion disk, which emits UV and soft X-ray radiation. However, in LLAGN, the truncated thin disk leads to a substantial reduction in these high-energy photons \citep[e.g.][]{Ho2008}, which in turn weakens [NeV] emission.

Figure~\ref{fig:nev_oiv} compares the luminosities of [NeV]$_{14 \mu \rm{m}}$ and [OIV]$_{26 \mu \rm{m}}$  with those reported in previous studies. Our sample includes the weakest [NeV]$_{14 \mu \rm{m}}$ and [OIV]$_{26 \mu \rm{m}}$ luminosities yet reported—those of M94—which are both detected at $>5\sigma$ significance. Despite their low luminosities, these lines follow the same correlation observed in higher-luminosity AGN, a relationship that spans seven orders of magnitude in luminosity.

We find a strong correlation between [NeV]$_{14\mu\rm{m}}$ and [OIV]$_{26\mu\rm{m}}$, in agreement with previous findings \citep[e.g.][]{Goulding2009}. This reinforces the view that these high–IP lines originate from the same emission and ionization regions. Other high–IP features, such as [MgIV]$_{4.49\mu\rm{m}}$ (IP = 80.4 eV), are detected in all but M87. Although underexplored, perhaps due to limited mid–IR coverage prior to JWST, [MgIV]$_{4.49\mu\rm{m}}$ displays similar linear luminosity trends with [NeV]$_{14\mu\rm{m}}$ and [O IV]$_{26\mu\rm{m}}$, but with greater scatter. We also detect [NeVI]$_{7.6\mu\rm{m}}$ (IP = 126 eV) in a subset of sources (NGC~1052, NGC~4258, NGC~4395, M~81, and Cen~A), adding further evidence for highly ionized gas. Given the ubiquity of [NeV]$_{14\mu\rm{m}}$ in our sample it serves as the most reliable benchmark for the comparative analysis that follows.

We use diagnostic line ratios between high- and low-IP emission lines to compare our observations with photoionization and shock model predictions in order to identify the dominant excitation mechanism. We also explore how these diagnostics vary with Eddington ratio to probe links between ionization and accretion state.

\subsubsection{Dominant Ionization Mechanism: Evidence for Photoionization}\label{sec:ionization_source}

\begin{figure*}
  \center
  \includegraphics[width=\linewidth]{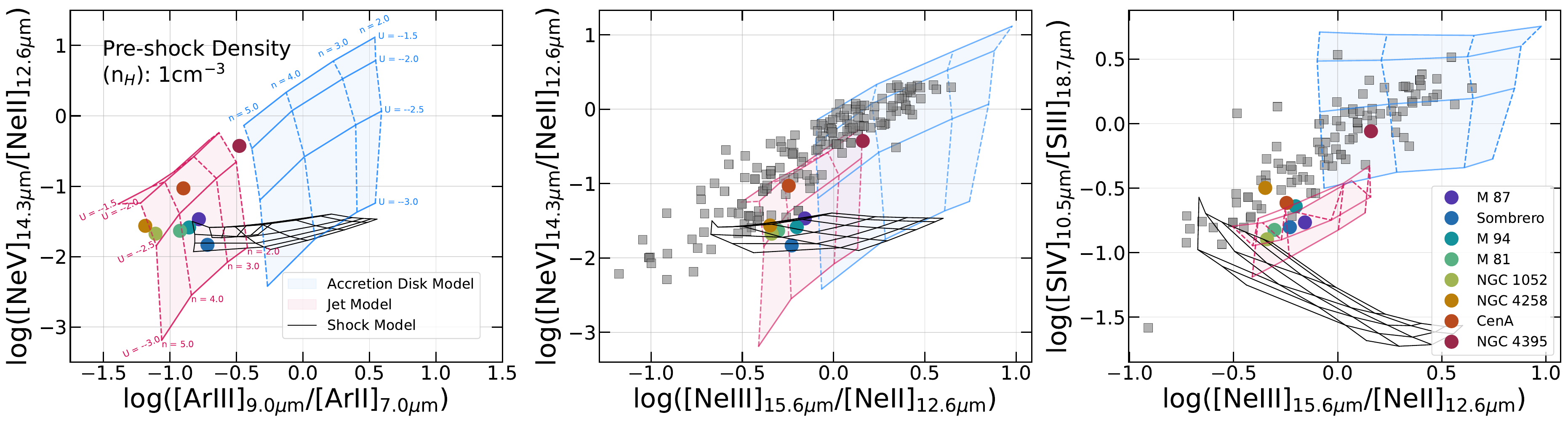}
  \caption{The emission line ratios observed in ReveaLLAGN targets are broadly consistent with photoionization models using a power-law ionizing spectrum (red grid; \citealt{Fern2023}). For NGC~4395, which has a relatively low black hole mass and accretes above a log Eddington ratio of –2.7, the line ratios are better reproduced by accretion disk–dominated photoionization models (blue grid; \citealt{Ferland2020}). Shock-only models (black grid) reproduce neon line ratios that are degenerate with those from power-law photoionization models, but only within a narrow range of low pre-shock densities and low shock velocities. At higher pre-shock densities, the shock models become inconsistent with the observations, and they fail to reproduce several key line ratios beyond neon. Gray squares represent comparison data as described in \ref{sec:supplemental_data}, and ReveaLLAGN sources are colored the same as in Figure~\ref{fig:nev_oiv}.}
  \label{fig:photo_vs_shock}
\end{figure*}

The detection of high-ionization lines such as [NeV]$_{14 \mu \rm{m}}$ and [NeVI]$_{10 \mu \rm{m}}$ in LLAGN raises the question of whether their origin lies in AGN photoionization or alternative mechanisms like fast radiative shocks. This distinction is particularly relevant in the low-Eddington regime, where the ionizing continuum may differ significantly from that of standard AGN. Observationally, both shocks and AGN can reproduce similar high-ionization signatures, but detailed line ratio diagnostics can help distinguish between them.

 In Figure~\ref{fig:photo_vs_shock}, we present a set of diagnostic diagrams to probe the ionization mechanisms at play. Each panel presents ratios that couple a mid- to high-ionization line (IP $\gtrapprox$ 23 eV) with a low-ionization line (IP $\lessapprox$ 23 eV). By comparing different such ratios, we probe variations in the ionizing spectrum and excitation conditions. The first plot shows [NeV]$_{14 \mu \rm{m}}$/[NeII]$_{12 \mu \rm{m}}$ (97 vs. 21 eV) vs. [ArIII]$_{9 \mu \rm{m}}$/[ArII]$_{7 \mu \rm{m}}$ (28 vs. 16 eV), the second plot shows [NeV]$_{14 \mu \rm{m}}$/[NeII]$_{12 \mu \rm{m}}$ (76 eV difference) vs. [NeIII]$_{15 \mu \rm{m}}$/[NeII]$_{12 \mu \rm{m}}$ (41 vs. 21 eV), and the last plot shows [SIV]$_{10 \mu \rm{m}}$/[SIII]$_{19 \mu \rm{m}}$ (35 vs. 23 eV) vs. [NeIII]$_{15 \mu \rm{m}}$/[NeII]$_{12 \mu \rm{m}}$. Ionization potentials for each emission line can also be found in Table~\ref{tab:line_results}. In each panel, we compare the observed emission-line ratios with theoretical predictions, overlaid as model grids representing collisional excitation from shocks and two photoionization models with differing ionizing SEDs. These plots allow us to evaluate whether the observed emission in our LLAGN sample is significantly driven by AGN photoionization or by shock excitation.

In all three plots, the black grid lines correspond to shock model predictions from the publicly available \href{http://3mdb.astro.unam.mx:3686/}{3MdBs database}, calculated using \texttt{MAPPINGS V} \citep{Sutherland2018} with elemental abundances from \citet{Gutkin2016}. These models assume a pre-shock hydrogen density of $n_{\rm H} = 1$~cm$^{-3}$ and a metallicity of $Z = 0.017$, with shock velocities ranging from 150 to 300~km~s$^{-1}$ and transverse magnetic fields from $10^{-4}$ to $1~\mu$G. We selected these parameter ranges because they provide the closest agreement with observed values of [NeV]$_{14 \mu \rm{m}}$/[NeII]$_{12 \mu \rm{m}}$ vs. [NeIII]$_{15 \mu \rm{m}}$/[NeII]$_{12 \mu \rm{m}}$ shown in the middle panel. We discuss models with other parameters more below. 

Alongside the shock models, we include two grids of photoionization models computed using \texttt{CLOUDY} v23.01 \citep{Chatzikos2023}. Both model sets show ionization parameters in the range log(U) = [–3, –1.5], and electron densities log(n$_{e}$/cm$^{3}$) = [2, 5]. 
The blue grid corresponds to disc models including a soft X-ray/UV bump component at intermediate Eddington accretion \citep{Jin2012}, representative of the ionizing continuum of moderate luminosity Seyferts and quasars \citep{Ferland2020}.
The red grid corresponds to models with an ionizing continuum characterized by a power-law from \citet{Fern2023}. The ionizing continuum is derived from the Sombrero galaxy SED, notable for lacking a thermal UV bump. This SED is characteristic of radiatively inefficient AGN, such as expected in LLAGN. A comprehensive description of the ionization models and the comparison with line ratios using high-ionization IR lines will be presented in a future work (Acharya et al., \textit{in prep}).

In the middle panel of Figure~\ref{fig:photo_vs_shock}, which compares neon line ratios, the shock models are degenerate with both the power-law and accretion disk-dominated photoionization models. However, this degeneracy is broken in the left and right panels, where neon ratios are compared to argon and sulfur ions. These diagrams reveal that while shock models can reproduce certain individual line ratios for some sources, they generally fail to match the full suite of observed diagnostics across the LLAGN sample. In particular, shock models over-predict [ArIII]$_{9 \mu \rm{m}}$/[ArII]$_{7 \mu \rm{m}}$ ratios and under-predict [SIV]$_{10 \mu \rm{m}}$/[SIII]$_{19 \mu \rm{m}}$ ratios, when compared with pure photoionization models.

The left panel highlights the diagnostic power of the [ArIII]$_{9 \mu \rm{m}}$/[ArII]$_{7 \mu \rm{m}}$ ratio, which provides strong separation between shock and photoionization models. Measurements for this ratio are absent from the archival datasets used in this study (see Section~\ref{sec:supplemental_data}) and are not reported in previous catalogs, making argon a relatively unexplored parameter space. It is a noble gas and like neon, doesn't deplete onto dust grains. Future mid-infrared spectroscopy with JWST, with its enhanced sensitivity and spectral coverage, could enable the use of argon-based diagnostics to place additional constraints on ionization mechanisms in LLAGN and other sources.

The inconsistency between shock model predictions and our data becomes more pronounced when assuming higher pre-shock gas densities, which would be expected in typical AGN \citep[e.g.][]{Feltre2023}.  This further disfavors shocks as the dominant ionization mechanism in our nuclear spectra. While contributions from shocked emission is likely present, especially in nuclei showing broad fine-structure lines indicative of outflows, it cannot account for the observed ratios alone. Instead, the emission from the ReveaLLAGN targets is consistently well reproduced across all three diagnostic diagrams by AGN photoionization models with a power-law ionizing continuum lacking a thermal UV bump. There are two exceptions to the general agreement with the power-law models: the first is NGC~4395, the highest Eddington ratio source in the sample, whose status as an LLAGN is due primarily to its low black hole mass \citep{Filippenko2003, Peterson2005}. The second is NGC~4258, also the second highest Eddington ratio source, which shows unusually large [SIV]$_{10 \mu \rm{m}}$ emission. Overall, the strong and systematic agreement between the observed line ratios and the power-law photoionization models indicates that the dominant excitation source in our LLAGN sample is a radiatively inefficient ionizing continuum without a thermal UV bump.

In addition to the mid-to-high ionization ratios shown in Figure~\ref{fig:photo_vs_shock} and discussed above, we also explored shock-sensitive diagnostics using only mid- to low-ionization lines. Specifically, we examined [ArIII]$_{9\mu{\rm m}}$/[ArII]$_{7\mu{\rm m}}$, [NeIII]$_{15\mu{\rm m}}$/[NeII]$_{12\mu{\rm m}}$, and [FeII]$_{5\mu{\rm m}}$/[FeII]$_{18\mu{\rm m}}$. Comparisons between these ratios show that shock models can reproduce the observed values in Sombrero, M87, and NGC~4258 for preshock densities between 10-100 cm$^{-3}$, shock velocities between 200-500 km s$^{-1}$, and magnetic field between 0.0001 and 3.23 $\mu$G.  All of these galaxies also show very broad lines, suggesting a role of outflows in these galaxies line emission. While, no single parameter set in the explored shock models provide a consistent match across multiple diagnostic ratios in any of our galaxies, these results do suggest that shocks are capable of contributing to the excitation and emission of low-IP lines in the nuclear region of LLAGN. This conclusion is consistent with the models presented in \citet{Contini1997} and \citet{Contini2001}, which explain LINER spectra as a combination of both photoionization and shock excitation.

\begin{figure}
  \center
  \includegraphics[width=\linewidth]{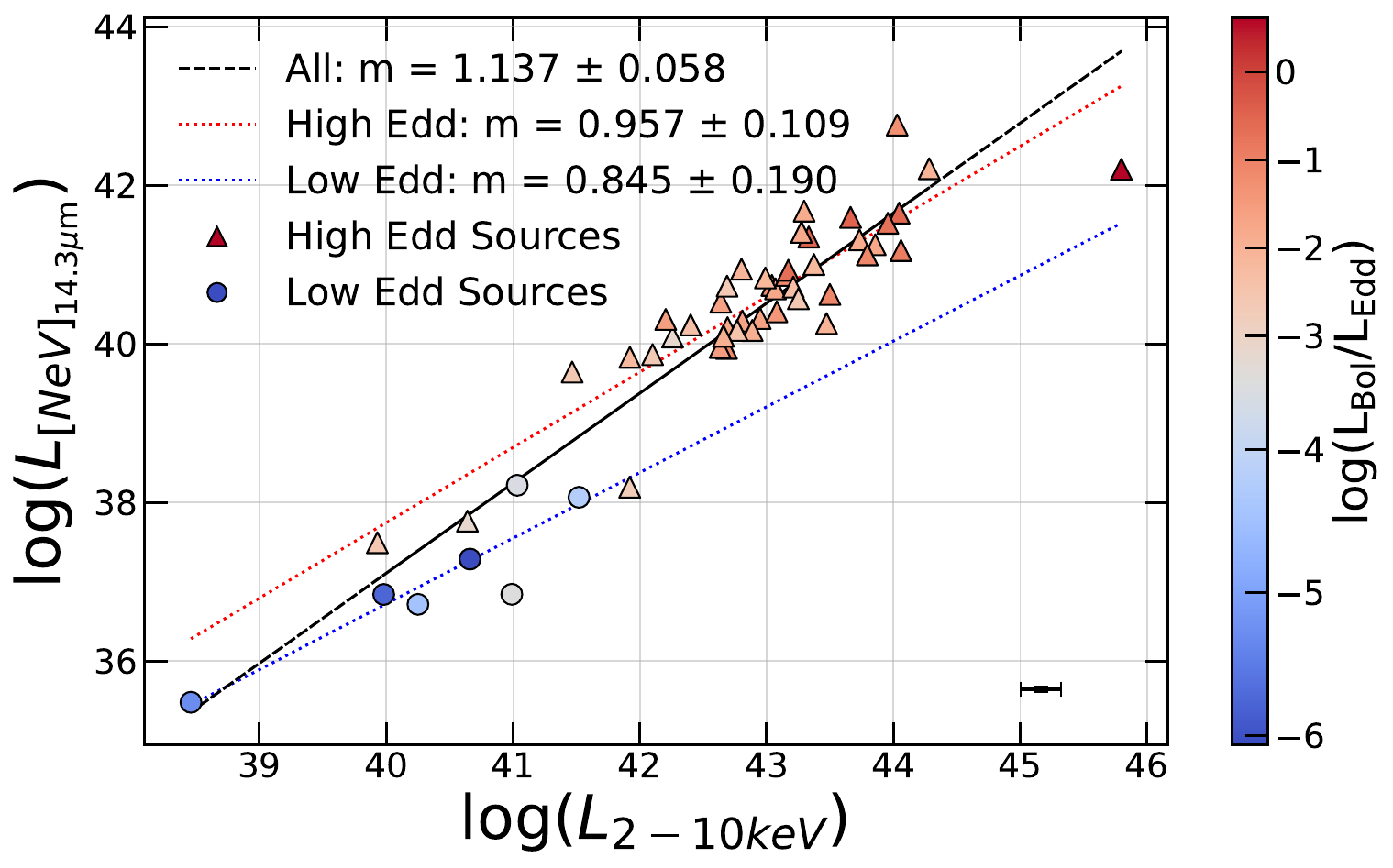}
  \caption{[NeV]$_{14 \mu \rm{m}}$ log luminosity versus 2–10 keV X-ray log luminosity in erg/s. Sources with a log(L$_{\rm{Bol}}$/L$_{\rm{Edd}}$) less than -3 are plotted as circles while sources with any higher Eddington ratios are plotted as triangles. ReveaLLAGN data is shown alongside data from \cite{Fernandez2016, Fernandez2021, Asmus2015} as discussed in Section~\ref{sec:supplemental_data}. The black dashed line shows a linear fit to all data. Red and blue dotted lines represent a conditional linear fit (Equation~\ref{eq:conditional_linear}) to the high and low Eddington sub-samples, respectively. BIC is lowest ($\Delta$BIC $\sim$9) when splitting the data at log(L$_{\rm{Bol}}$/L$_{\rm{Edd}}$) 
  = -3.5. Marker colors indicate log(L$_{\rm{Bol}}$/L$_{\rm{Edd}}$), with gray set to match the best-fit dividing log(L$_{\rm{Bol}}$/L$_{\rm{Edd}}$) $\sim$ -3.5. Characteristic 1$\sigma$ errors are shown in the bottom right of the plot.} 
  \label{fig:nev_lx}
\end{figure}

\subsubsection{Eddington Ratio Dependence}\label{sec:edd_dependence}

This section examines how IR line emission and diagnostic line ratios vary with the Eddington ratio of the central AGN, a parameter widely expected to mark changes in the structure of the central engine \citep[e.g.][]{Narayan1995}. Our ReveaLLAGN emission line sample probes not just lower luminosities, but also lower Eddington ratios than previously possible. 

Due to its AGN origin, the [NeV]$_{14\mu \rm{m}}$ line has been used to estimate the bolometric luminosity of AGN \citep{Satyapal2007, Satyapal2008}. Likewise, the 2--10 keV X-ray luminosity is a well established and widely adopted indicator of bolometric luminosity \citep{Duras2020}. High-IP mid-IR lines, including [OIV]$_{26\mu \rm{m}}$ and [NeV]$_{14\mu \rm{m}}$, have also been found to correlate with the soft (0.5--2 keV) X-ray emission in luminous AGN samples \citep{Prieto2002, RodriguezArdila2011}, pointing to a general link between coronal lines and the ionizing continuum. Extending this comparison, we examine the relation between [NeV]$_{14\mu \rm{m}}$ and the 2--10 keV luminosity and find a strong, nearly linear, correlation (slope = 1.137$\pm$0.058) shown in Figure~\ref{fig:nev_lx}. However, the lowest luminosity and lowest Eddington ratio sources, including most galaxies in our sample, systematically fall below this relation. We therefore also apply a conditional linear model where the slope and intercept depend on an Eddington ratio threshold: 
\begin{equation}\label{eq:conditional_linear}
y =
\begin{cases}
  m_1 x + k_1, & \text{if } z < z_0 \\
  m_2 x + k_2, & \text{if } z \ge z_0.
\end{cases}
\end{equation}
 The best-fit dividing Eddington ratio was found to be at log$(L_{\mathrm{bol}}/L_{\mathrm{Edd}}) = -3.5^{+0.6}_{-1.0}$. This conditional linear model (BIC: -66.42) outperforms the simple linear model fit to the full data set (BIC: -57.08). This suggests the Eddington ratio cutoff, which is an empirical transition point in the L$_X$ to L$_{NeV}$ relation, may correspond to a transition in the ionizing structure of the AGN. The lines representing high- and low-Eddington bins are parallel with sublinear slopes of $0.957 \pm0.109$ and $0.845 \pm0.190$, but vertically offset, indicating a systematic reduction in the [NeV] output in low-Eddington sources. The observed offset between high- and low-Eddington sources is larger than the $\sim 0.1$~dex variation expected from the spread in X-ray bolometric corrections ($k_X \sim 16$--20; see Section~\ref{sec:bolometric corrections}, indicating a genuine decrease in [NeV] output. This decrease in [NeV] emission is nearly an order of magnitude for sources below log($L_{\mathrm{bol}}/L_{\mathrm{Edd}}) \approx$ –3.5.

This interpretation is reinforced by re-examination of the middle panel of Figure~\ref{fig:photo_vs_shock}. While more luminous AGN trace a well-known sequence from star formation-contaminated to pure AGN-dominated spectra, our low-Eddington sources diverge from this trend. These ReveaLLAGN sources maintain relatively typical [NeIII]$_{15 \mu \rm{m}}$/[NeII]$_{12 \mu \rm{m}}$ ratios but show lower [NeV]$_{14 \mu \rm{m}}$/[NeII]$_{12 \mu \rm{m}}$, placing them in a sparsely populated region of parameter space. This pattern suggests that low-Eddington sources produce relatively less [NeV] emission than more luminous AGN.

\begin{figure}
  \center
  \includegraphics[width=\linewidth]{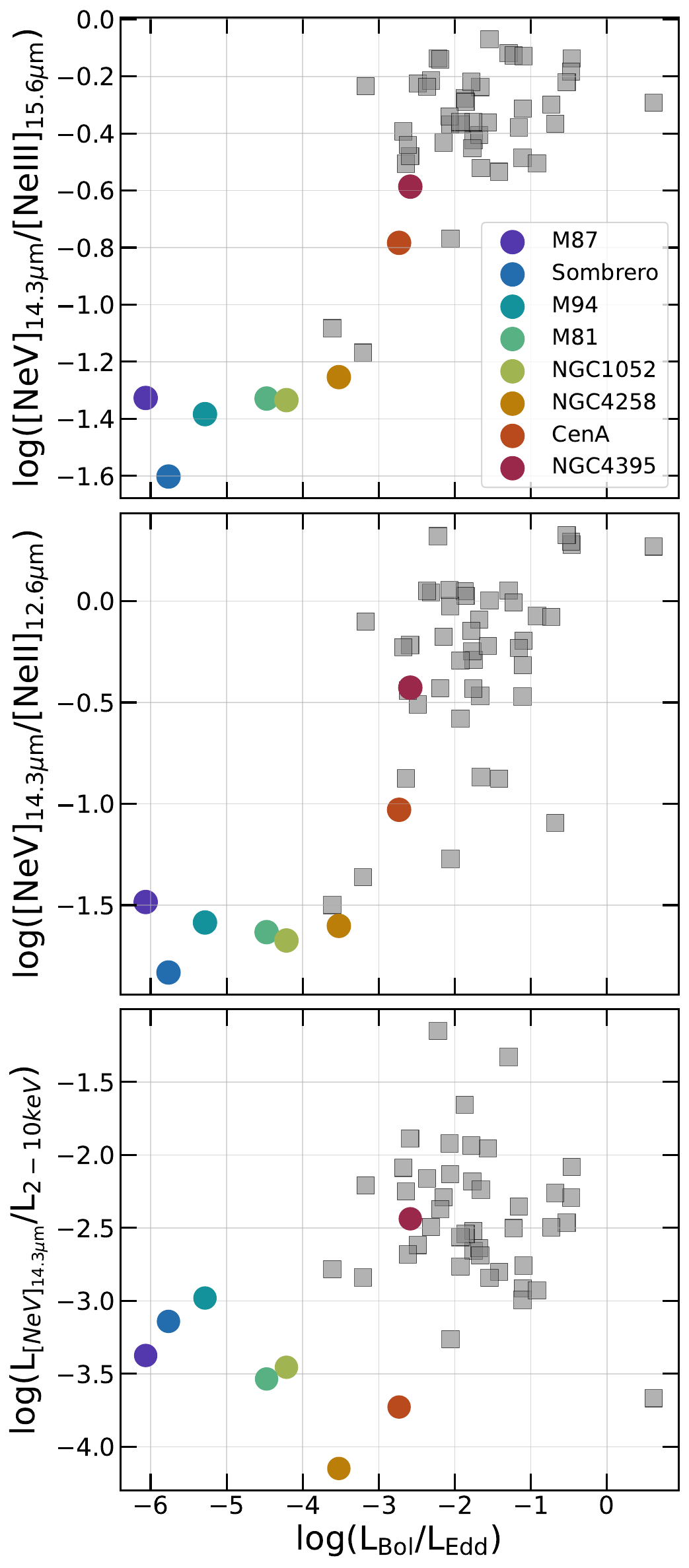}
  \caption{[NeV] ratios are correlated with Eddington Ratio. The x axis represents Eddington ratio ($\log(L_{\mathrm{bol}}/L_{\mathrm{Edd}}$). The top plot compares log([NeV]$_{14 \mu \rm{m}}$/[NeIII]$_{15 \mu \rm{m}}$), the middle plot compares log([NeV]$_{14 \mu \rm{m}}$/[NeII]$_{12 \mu \rm{m}}$), and the bottom plot compares log(L$_{[\rm{NeV}]}$/L$_{2-10\rm{keV}}$). In each plot gray squares represent data compiled from \citet{ Asmus2015, Fernandez2016, Fernandez2021}, while the colored circles represent ReveaLLAGN data, colors are the same as in \ref{fig:nev_oiv}.}
  \label{fig:ne_edd}
\end{figure}

We further explore this behavior in Figure~\ref{fig:ne_edd}, which explicitly examines [NeV]$_{14 \mu \rm{m}}$ line ratios as a function of Eddington ratio. The top two panels show [NeV]$_{14 \mu \rm{m}}$/[NeII]$_{12 \mu \rm{m}}$ and [NeV]$_{14 \mu \rm{m}}$/[NeIII]$_{15 \mu \rm{m}}$ both sensitive to the hardness of the ionizing continuum. The [NeII]$_{12 \mu \rm{m}}$ line (ionization potential 21.6 eV) can arise from both star formation and AGN activity, while [NeIII]$_{15 \mu \rm{m}}$ (41.0 eV) requires a harder radiation field. Gray squares denote a comparison sample of well-characterized AGN from the literature (see Section~\ref{sec:supplemental_data}), while colored circles represent ReveaLLAGN sources.

ReveaLLAGN sources display significantly lower [NeV]$_{14 \mu \rm{m}}$/[NeII]$_{12 \mu \rm{m}}$ and [NeV]$_{14 \mu \rm{m}}$/[NeIII]$_{15 \mu \rm{m}}$ ratios than higher Eddington sources, noticeable at $\log(L_{\mathrm{bol}}/L_{\mathrm{Edd}}) \sim -3$ effectively separating the low- and high- Eddington ratio populations. This division mirrors the cutoff found in the conditional linear function (Equation~\ref{eq:conditional_linear}) shown in Figure~\ref{fig:nev_lx}.

The same trend appears in the bottom panel of Figure~\ref{fig:ne_edd}, which compares L$_{[NeV]_{14 \mu \rm{m}}}$/L$_{X\,(2-10~\mathrm{keV})}$ with Eddington ratio. Here too, a clear drop in relative [NeV] strength is seen at $\log(L_{\mathrm{bol}}/L_{\mathrm{Edd}}) \lesssim -3$, reinforcing the view that high-energy ionizing photon output diminishes near this accretion rate.

\subsection{Evidence of Kinetic Mode Feedback on Molecular Gas}\label{sec:kin_mode_feedback_discussion}

Low-Luminosity Active Galactic Nuclei represent a mode of black hole accretion characterized by radiative inefficiency and the dominance of mechanical feedback through compact jets and outflowing winds \citep[e.g.][]{Trump2011}. This kinetic-mode feedback injects energy and momentum into the surrounding molecular gas, impacting star-formation and galaxy evolution \citep{Ho2008, Yuan2014, Heckman2014, Mukherjee2018}. In this section we explore two diagnostics that are linked to kinetic feedback in galaxies; the excitation temperature of molecular hydrogen, and the ratio of molecular hydrogen to PAH emission.

\begin{figure}
  \center
  \includegraphics[width=\linewidth]{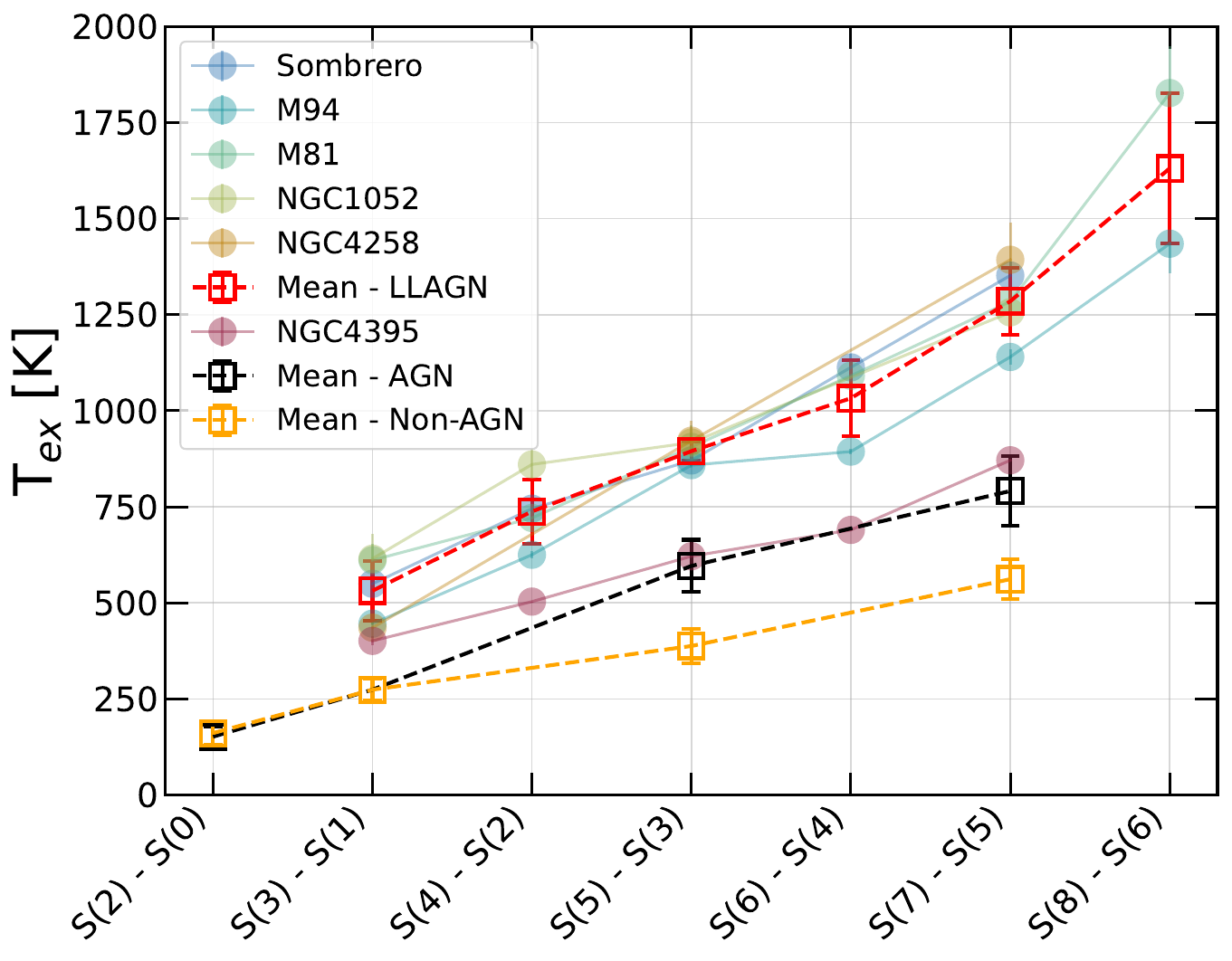}
  \caption{Molecular hydrogen in LLAGN targets from the ReveaLLAGN survey exhibits systematically higher excitation temperatures compared to AGN and non-AGN galaxies. This temperature difference increases with higher rotational transitions. We present a plot of pairwise excitation temperatures between consecutive rotational transitions of the same parity. The mean excitation temperatures for LLAGN are shown in red, while the corresponding mean values for AGN and non-AGN galaxies, adopted from \cite{Lambrides2019}, are shown in black and orange.}
  \label{fig:h2_temp}
\end{figure}

In our sample of LLAGN, we detect high signal-to-noise pure rotational H$_2$ lines in all galaxies except M87 and Cen~A (Table~\ref{tab:number_of_emission_lines}). We calculate excitation temperatures using line pairs of the same parity (Section~\ref{sec:measure_h2_excitation}) and in Figure~\ref{fig:h2_temp} we compare the mean excitation temperatures for each line pair in the ReveaLLAGN sample to the mean temperatures of the AGN and non-AGN populations from \citet{Lambrides2019}. In that study, galaxies are classified as AGN-dominated (black line) if the equivalent width (EW) of the 6.2~$\mu$m PAH feature is $<$ 0.27~$\mu$m and non-AGN (orange line) galaxies have an EW $>$ 0.27~$\mu$m.

In our sample, excitation temperatures derived from pure rotational (0-0 S(1-8)) transitions of molecular hydrogen are systematically higher than in the comparison samples of both AGN- and non-AGN-dominated galaxies from \citet{Lambrides2019} (Figure~\ref{fig:h2_temp}). The notable exception is NGC~4395, our highest Eddington ratio galaxy, which shows excitation temperatures consistent with the AGN population. The lowest transitions (S(1)--S(3)) show the smallest offset from the comparison sources. These lower transitions are primarily heated by star formation activity in both AGN and Non-AGN sources \citep{Rigopoulou2002, Higdon2006, Roussel2007}. At higher transitions this difference grows, the mean excitation temperature for S(7)-S(5) in LLAGN is ~500K higher than the mean in the AGN population. These results suggest that LLAGN host an additional or enhanced heating channel for warm molecular gas beyond the mechanisms typical of luminous AGN and star-forming systems. 

In sources with independent evidence for jets or outflows, shocks are a natural explanation for the elevated excitation temperatures. Jet/ISM and wind/ISM interactions transfer kinetic energy into molecular clouds, producing elevated excitation temperatures \citep[e.g.,][]{Ogle2010, Riffel2013, Riffel2020, Garcia2024}. The observed excitation levels in the near and mid IR range available to JWST further support this picture: nearly all LLAGN nuclei show many strong 0--0 rotational lines, few-to-no 1--0 rovibrational lines, and essentially no 2--1 lines. This combination of strong rotational lines with little rovibrational features is characteristic of C-type shocks \citep{Kristensen2023}, which efficiently populate low- and mid-$J$ rotational levels while producing relatively little rovibrational emission. Another heating mechanism that populates the same rotational levels is found in X-ray dominated regions (XDRs), where hard X-rays penetrate deeply into molecular gas and efficiently raise its temperature. \citet{Ogle2010} find that X-ray heating is energetically viable when $L_{\rm H_2}/L_{2-10,{\rm keV}} < 0.01$. We calculate $L_{\rm H_2}$ from all available S(0)--S(8) lines and find that all nuclei in our sample, except M94, fall below this threshold, indicating that our observations are consistent with X-ray heating as a viable mechanism.

To further assess excitation mechanisms, we compute H$_2$ S(3)/PAH$_{11.3\mu\mathrm{m}}$ ratios \citep{Lambrides2019, Riffel2025}. The H$_2$/PAH ratio is a sensitive diagnostic of the dominant heating process because PAH emission primarily traces UV excitation from young stars, whereas warm H$_2$ can also be excited by shocks or X-rays \citep[e.g.,][]{Roussel2007, Ogle2010}. In environments where shocks or outflows are present, these processes not only heat molecular gas but can also destroy small PAH grains, leading to suppressed PAH emission and elevated H$_2$/PAH ratios \citep{Riffel2020, Zhang2022}. The 11.3~$\mu$m PAH feature is detected at $>3\sigma$ in all targets but M87 (see Section~\ref{sec:measure_nuclear_pah} and Section~\ref{sec:appendix_PAH}). We present this in Table~\ref{tab:pah_sil_params}. Our LLAGN sample has a median H$2$/PAH${11.3\mu\mathrm{m}}$ ratio of 0.044 with a standard deviation of 0.207, excluding M87 and Cen~A where this ratio is unconstrained. The spread is driven largely by NGC~4395 which has the largest H$_2$/PAH$_{11.3\mu\mathrm{m}}$ ratio of 0.642. These values are consistent with, and in some cases exceed, the median AGN ratio of 0.055, indicating an H$_2$ excess comparable to the AGN population from \citet{Lambrides2019}. In NGC~4395, the high ratio and weak PAH emission may also reflect its low metallicity (12 + log O/H $\sim$ 8.3; \citealt{Roy1996}), since PAHs form less efficiently and are more easily destroyed in metal-poor environments \citep{Engelbracht2005, Wu2006, O'Halloran2006, Draine2007B, Sandstrom2012}.  

These pure rotational diagnostics suggest that LLAGN contain warmer molecular gas than is typical for either luminous AGN or star-forming galaxies. Shocks from jets and outflows in some combination with XDR heating could be responsible for the observed warm H$_2$ budget in our sources. Future progress will require spatially resolved maps with kinematic diagnostics to disentangle regions dominated by shocks and XDRs.

\subsection{Broad Silicate Emission}\label{sec:silicate_discussion}

\begin{figure}
  \center
  \includegraphics[width=\linewidth]{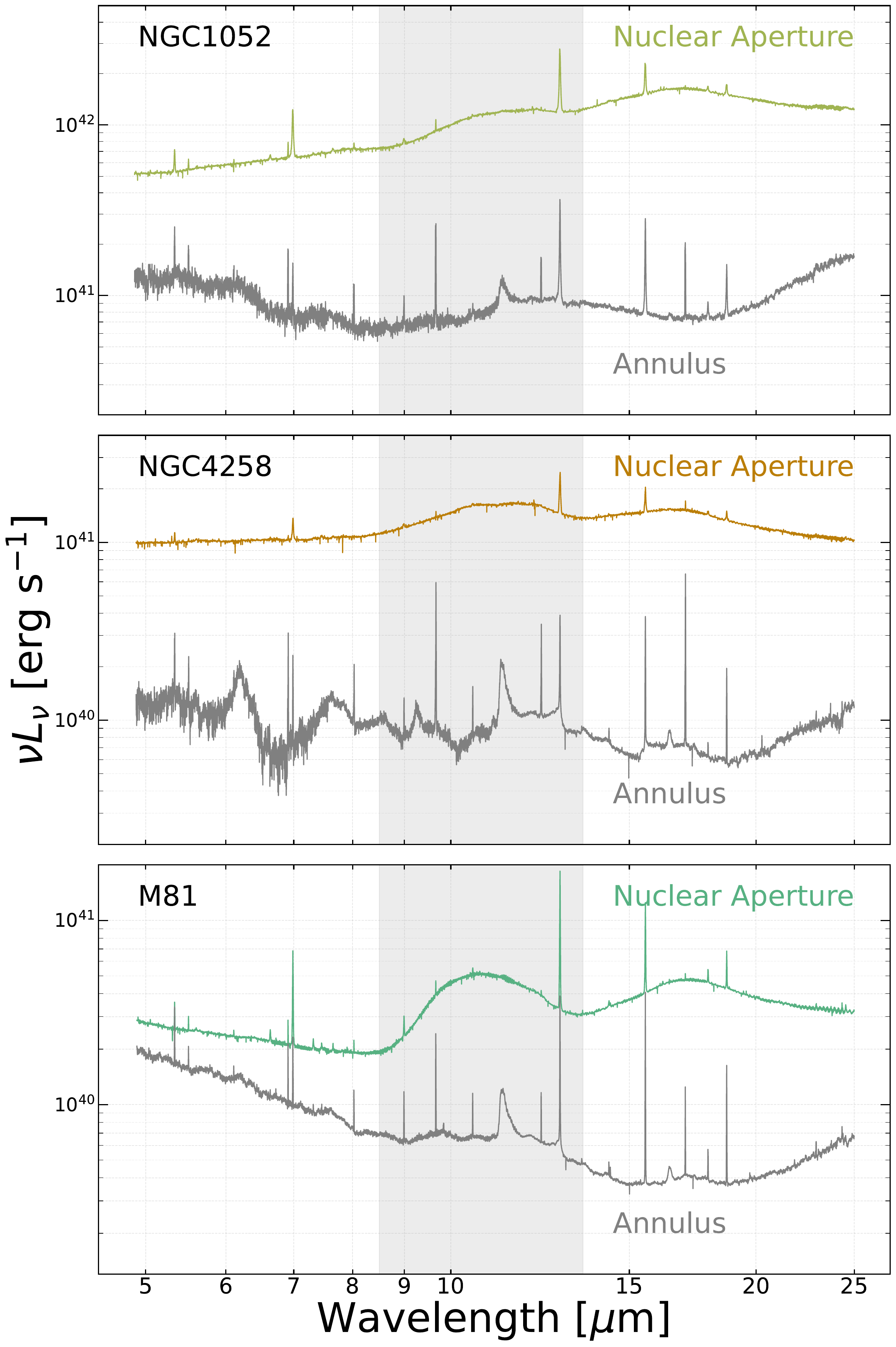}
  \caption{Broad silicate emission, peaking at 10.5 $\mu$m (shaded gray), is seen in the nuclear extracted spectra of most of our target galaxies (see Table~\ref{tab:pah_sil_params}, but is weak or absent in annular spectra. We show three galaxies; NGC~1052, NGC~4258, and M81, to highlight this trend. In each plot, the nuclear extracted spectra from Section~\ref{sec:nuclear_extraction} is shown in color, while the gray spectra is extracted from an annulus with r$_{in}$ = 1.5\arcsec and r$_{out}$ = 2\arcsec.}
  \label{fig:silicate_emission}
\end{figure}

Broad silicate features at 9.7 and 18~$\mu$m are prominent MIR spectral signatures arising from Si--O stretching and O--Si--O bending modes in amorphous silicate dust grains. Whether these features appear in absorption or emission can depend on the geometry, distribution, and optical depth of the surrounding dust \citep{Pier1992, Granato1994, Efstathiou1995}. Absorption occurs when cool silicate dust obscures the central source along the line of sight, typically in edge-on systems such as Type~2 AGN, in heavily embedded ``buried'' nuclei, or from dust in the host galaxy due to dust lanes or disturbed morphologies \citep{Goulding2012}. In contrast, silicate emission requires optically thin dust and occurs when warm dust is directly illuminated by the AGN and re-emits thermally in the MIR, as is expected in the case in Type~1 AGN viewed face-on. However, Type~1 AGN are equally found with both silicate emission and absorption \citep{Hao2007}, and some Type~2 AGN, quasars, and LINERS have been observed with silicate emission of varying strength \citep{Sturm2006, Mason2009, Nikutta2009, Hatziminaoglou2015}. 

The silicate Si–O stretching feature in the Milky Way's diffuse ISM typically peaks at $\sim$9.7 $\mu$m \citep{Kemper2004, Chiar2006, Henning2010}. However, in many AGN, the 9.7 $\mu$m silicate emission is broadened and red-shifted $>$ 10 $\mu$m, inconsistent with standard Galactic silicates. Grain size, porosity, shape, composition, and temperature can all contribute in shifting and broadening silicate emission \citep{sturm2005, li2008, Smith2010, Xie2017}. Large, porous grains can form through coagulation after being destroyed then reconstituted via AGN activity \citep{Smith2010}.

Table~\ref{tab:pah_sil_params} presents the silicate strength measurements (Section~\ref{sec:measure_nuclear_silicate}) for each galaxy. Five of the seven ReveaLLAGN targets exhibit clear silicate emission, while M87 and NGC~4395 show no clear silicate emission or absorption. In contrast, the nuclear spectrum of Cen~A displays strong silicate absorption due to the prominent foreground dust lane in this galaxy.

To investigate the spatial scale of the silicate emission, we compare nuclear-extracted spectra to spectra extracted from annular regions with r$_{in}$ = 1.5\arcsec\ and r$_{out}$ = 2.0\arcsec\ for NGC~1052, NGC~4258, and M81 (Figure~\ref{fig:silicate_emission}). While the nuclear aperture varies with wavelength (Section~\ref{sec:nuclear_extraction}), its size at the 10.5 $\mu$m silicate peak is 0.4\arcsec, corresponding to physical scales of 35 pc, 14 pc, and 7 pc for NGC~1052, NGC~4258, and M81, respectively. The annular region spans 141–188 pc for NGC~1052, 55–74 pc for NGC~4258, and 27–36 pc for M81.

The annulus was chosen to provide a well-defined off-nuclear sampling region while remaining within empirically supported limits. AGN mid-IR continuum emission is typically compact (FWHM $\lesssim$ 1\arcsec; \citealt{Asmus2019}) but can extend to $\sim$2.5\arcsec\ with JWST/MIRI \citep{Leist2024}. The 1.5\arcsec\ inner edge lies well outside the MIRI PSF FWHM \citep{Argyriou23}, while 2.0\arcsec\ is the maximum radius accessible across all MIRI/MRS channel cubes. 

Broad silicate emission features at 10 $\mu$m are clearly visible in the nuclear spectra but appear much weaker, or are entirely absent, in the larger annular extractions. Further analysis of the spatial profile (Ohlson et al., {\em in prep}) indicates that the silicate emission is mostly unresolved. This suggests that the emission originates on parsec scales near the black hole. To interpret the silicate emission observed in our LLAGN sample we consider several possible origins that have been proposed in the literature.

Dust temperatures of $\sim$200--300~K are inferred from modeling silicate-emitting regions \citep{Xie2017, Smith2010}, much cooler than the sublimation temperature ($\sim$1500~K) near the hot inner edge of the torus \citep{Sturm2006}. This temperature aligns more closely with that of extended dust in the narrow-line region (NLR), where mid-infrared imaging shows the presence of NLR dust at $\sim$200–300 K, extending 1–2\arcsec\ from AGN nuclei \citep{Bock2000, Radomski2003, Gorjian2004, Packham2005}. Dust heated by the AGN but not part of the torus, could explain silicate emission seen in some Type~2 AGN. The proximity of our targets and the point-like nature of the Silicate emission means that the emission we observe must be coming from pc-scale spatial scales.

Clumpy torus models allow for cool dust near the nucleus and can produce silicate emission even along equatorial sight lines \citep{Nenkova2008}, though not always as strong as observed in LLAGN \citep{Mason2013}. Some models suggest that the dusty torus may disappear in LLAGN, either due to insufficient inflow of material from the host galaxy or because disk-driven winds, which sustain the torus, cease to operate at low accretion rates \citep{Elitzur2006}. Without a standard Seyfert-like torus, the silicate emission in LLAGN could instead originate from diffuse, optically thin dust, representing the remnants of a dissipating torus structure \citep{Mason2013}. 

Silicate emission could also arise from parsec-scale dust filaments in the nuclear region. In Sombrero, such filaments are optically thin at the nucleus but become optically thicker several parsecs away \citep{Prieto2014}, while in NGC~1052 a single filament crosses immediately adjacent to the nucleus \citep{Prieto2021}. Both are consistent with strong silicate emission in compact nuclear extractions but weaker features in annular apertures. In contrast, M87 shows little to no central dust \citep{Prieto2021}, consistent with our finding of little to no silicate emission.

\section{Conclusions}\label{sec:conclusion}
JWST spectroscopy from the ReveaLLAGN survey of NGC~1052, NGC~4258, NGC~4395, M81, M87, M94, and Sombrero, combined with archival JWST data of Cen~A, reveals several key insights into the ionization structure, molecular gas properties, and nuclear dust environment of LLAGN:

\begin{itemize}
    \item \textbf{Rich Emission-Line Spectra:}
    We create a linelist of 131 emission lines, with the dwarf galaxy NGC 4395 displaying the richest spectrum (106 detected lines) and M87 the sparsest (15 detected lines). High-ionization features such as [NeVI]${7.6 \mu \rm{m}}$, [NeV]${14.3 \mu \rm{m}}$, [OIV]${26 \mu \rm{m}}$, [MgIV]${4.49 \mu \rm{m}}$, and [ArVI]$_{4.53 \mu \rm{m}}$ are present in nearly every target, several detected for the first time in LLAGN, improving upon previous upper limits. Their concentration within sub-arcsecond apertures demonstrates that JWST can isolate nuclear emission with minimal host-galaxy contamination, allowing us to probe physical conditions and the faint ionizing output of LLAGN on scales of only a few parsecs in the nearest targets.

    \item \textbf{Accretion-Dependent Ionizing Continuum:} 
    We utilize high-ionization-potential (high-IP) emission lines and diagnostic line ratios to probe the ionizing continuum of LLAGN. To quantify the relationship between [NeV]$_{14.3 \mu \rm{m}}$ emission and absorption-corrected 2--10~keV X-ray luminosity, we empirically fit the data using a conditional linear model, which allows for separate linear fits in distinct regimes of Eddington ratio. This approach effectively bins the data based on $\log(L_{\rm bol}/L_{\rm Edd})$  above and below a critical threshold. The fit reveals a clear transition in [NeV]$_{14.3 \mu \rm{m}}$ luminosity at $\log(L_{\rm bol}/L_{\rm Edd}) = -3.5^{+0.8}_{-0.7}$. Below this threshold, [NeV]$_{14.3 \mu \rm{m}}$ emission is systematically weaker by up to an order of magnitude, indicating a significant deficit of ultraviolet ionizing photons.
    
    In addition, diagnostic line ratios such as [NeV]$_{14 \mu \rm{m}}$/[NeII]$_{12 \mu \rm{m}}$, [ArIII]$_{9 \mu \rm{m}}$/[ArII]$_{7 \mu \rm{m}}$, and [SIV]$_{10 \mu \rm{m}}$/[SIII]$_{19 \mu \rm{m}}$ show that LLAGN occupy a region of ionization parameter space distinct from that of luminous AGN and are consistent with an ionizing continuum that lacks a strong ultraviolet bump. Together, the empirical weakening of [NeV]$_{14.3 \mu \rm{m}}$ and the distinct ionization ratios point to a transition from a standard thin accretion disk to a radiatively inefficient accretion flow.
    
    Of particular note, the previously underutilized [ArIII]$_{9 \mu \rm{m}}$/[ArII]$_{7 \mu \rm{m}}$ ratio provides a sensitive diagnostic capable of distinguishing both between different photoionization models and between photoionization and shock-dominated emission.
    \item \textbf{Enhanced Molecular Gas Excitation:}
    LLAGN in our sample exhibit systematically elevated H$_2$ rotational temperatures, with excitation temperatures $\sim$500~K higher than those in luminous AGN and substantially above those in star-forming galaxies. These elevated temperatures are accompanied by high H$2$/PAH${11.3,\mu\mathrm{m}}$ ratios—comparable to, or exceeding, the AGN population median—indicating non-stellar heating and PAH suppression in their nuclei. The combination of warm molecular gas and weak PAH emission points to mechanical and radiative feedback, where shocks from compact jets or outflows and X-ray heating from the central engine both contribute to the excitation of molecular gas.
    
    \item \textbf{Nuclear Dust Properties:} 
    Broad silicate emission features peaking at 10.5~$\mu$m are detected in nearly all sources, with the exceptions of M87, which shows no silicate emission, and Cen~A, which exhibits silicate absorption. The silicate emission is consistent with being unresolved, suggesting its origin is optically thin dust localized near the central engine.

  
\end{itemize}

These findings collectively highlight the differences between LLAGN and their higher-luminosity counterparts, emphasizing the structural complexity, distinct ionizing conditions, feedback mechanisms, and nuclear environments that characterize these low-power systems.

\begin{acknowledgments}

We would like to thank Katja Fahrion for sharing MIRI/MRS and NIRSpec data on Cen~A. KG, AS, and DO acknowledge support from JWST Cycle 1 grant GO-2016. We acknowledge the ERO team for developing their observing program with a zero-exclusive-access period. LCH was supported by the National Science Foundation of China (12233001), the National Key R\&D Program of China (2022YFF0503401), and the China Manned Space Program (CMS-CSST-2025-A09).
 
 \end{acknowledgments}

\vspace{5mm}
\facilities{JWST (MIRI/MRS, NIRSPec)}

\software{astropy \citep{astropy2022}, lmfit \citep{lmfit2024}, jwst calibration pipeline v1.15.0 \citep{jwstpipeline}}

\appendix
\section{PAH 11.3 \texorpdfstring{$\mu$\lowercase{m}}{mu} -- Example Fits}\label{sec:appendix_PAH}

\begin{figure}[ht]
  \centering
  \includegraphics[width=\textwidth]{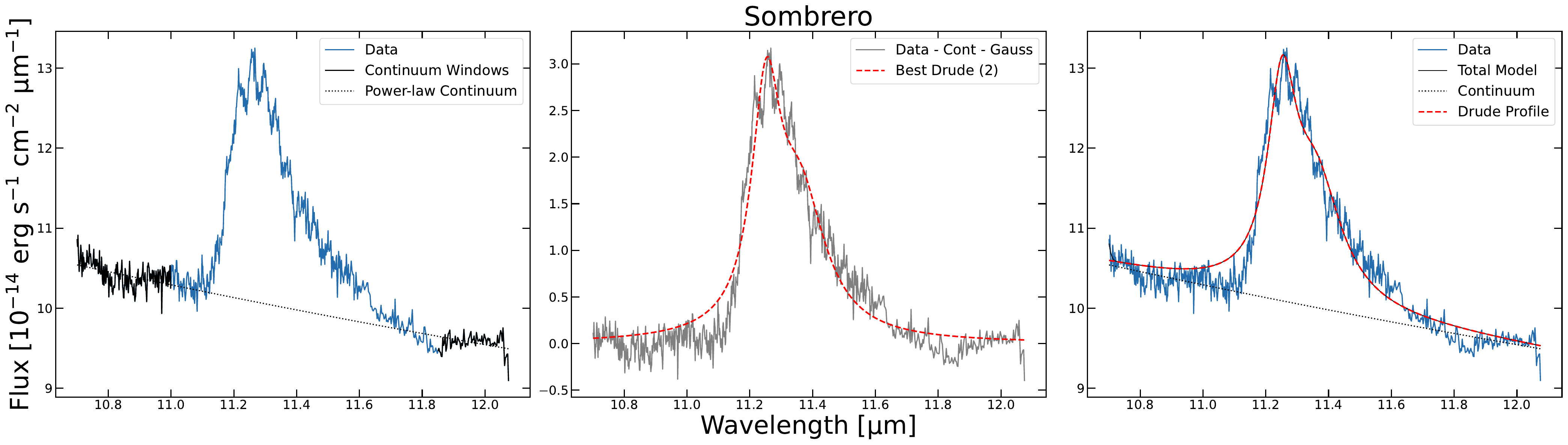}
  \caption{Fitting PAH 11.3 $\mu$m. \textit{Left Panel} -- shows the  power-law fit to the continuum on either side of the PAH feature. \textit{Middle Panel} -- shows the continuum subtracted data and a dashed red line representing the best multi-drude profile. This profile is used to measure flux and luminosity. \textit{Right Panel} -- shows the PAH model + continuum. }
  \label{fig:fitting_pah}
\end{figure}

\section{Emission Line Measurements}\label{sec:appendix_table}

\movetabledown=6.5cm
\begin{rotatetable}
\begin{deluxetable*}{ccccccccccccccc}
\tabletypesize{\footnotesize}
\setlength{\tabcolsep}{0.03in}
\tablewidth{0pt}
\tablecaption{Nuclear Spectra Emission Line Measurements \label{tab:line_results}}
\tablehead{
\colhead{Galaxy} & \colhead{Line} & \colhead{Wavelength\tablenotemark{a}} & \colhead{IP\tablenotemark{b}} & \colhead{Transition} & \colhead{Flux} & \colhead{Flux UL} & \colhead{Flux Err} & \colhead{Peak Vel} & \colhead{Peak Vel Err} &\colhead{ FWHM$_{\rm line}$} & \colhead{FWHM$_{\rm line}$ Err} & \colhead{S/N} & \colhead{Aperture Radius} & \colhead{Warning}\\
\colhead{ } & \colhead{ } & \colhead{$\mathrm{\mu m}$} & \colhead{$\mathrm{eV}$} & \colhead{} &\colhead{$10^{-14}\mathrm{erg\,s^{-1}\,cm^{-2}}$} &\colhead{$10^{-14}\mathrm{erg\,s^{-1}\,cm^{-2}}$} & \colhead{$10^{-14}\mathrm{erg\,s^{-1}\,cm^{-2}}$ } & \colhead{$\mathrm{km\,s^{-1}}$} & \colhead{$\mathrm{km\,s^{-1}}$} & \colhead{$\mathrm{km\,s^{-1}}$} & \colhead{$\mathrm{km\,s^{-1}}$} & \colhead{} &\colhead{pc} & \colhead{}} 
\startdata
NGC4395 & [Fe II] & 1.664 & 7.90 & -- &--& 0.003 &--&--&--&--&--&--& 3 & 7 \\
NGC4395 & [Fe II] & 1.677 & 7.90 & -- &--&--&--&--&--&--&--&--& 3 & 3 \\
NGC4395 & H2 & 1.680 & 0.00 & (2-0)O(9) &--&--&--&--&--&--&--&--& 3 & 3 \\
NGC4395 & H & 1.681 & 0.00 & 11-4 &--&--&--&--&--&--&--&--& 3 & 3 \\
NGC4395 & H2 & 1.688 & 0.00 & (1-0)S(9) & 0.010 &--& 0.002 & 80 & 30 & 380 & 100 & 5.9 & 3 & 0 \\
NGC4395 & H2 & 1.715 & 0.00 & (1-0)S(8) &--& 0.001 &--&--&--&--&--&--& 3 & 7 \\
NGC4395 & H & 1.737 & 0.00 & 10-4 & 0.047 &--& 0.002 & -10 & 10 & 220 & 20 & 14.4 & 3 & 1 \\
NGC4395 & [Fe II] & 1.745 & 7.90 & -- &--& 0.000 &--&--&--&--&--&--& 3 & 2 \\
NGC4395 & H2 & 1.748 & 0.00 & (1-0)S(7) &--& 0.000 &--&--&--&--&--&--& 3 & 2 \\
NGC4395 & H2 & 1.788 & 0.00 & (1-0)S(6) &--& 0.002 & 0.001 &--&--&--&--& 2.4 & 3 & 0 \\
NGC4395 & [Fe II] & 1.810 & 7.90 & -- & 0.020 &--& 0.001 & -20 & 10 & 130 & 10 & 23.8 & 3 & 0 \\
NGC4395 & H & 1.818 & 0.00 & 9-4 & 0.034 &--& 0.003 & -10 & 10 & 150 & 20 & 12.7 & 3 & 0 \\
NGC4395 & H2 & 1.836 & 0.00 & (1-0)S(5) & 0.013 &--& 0.001 & -20 & 10 & 70 & 10 & 23.1 & 3 & 0 \\
NGC4395 & [Cr II] & 1.841 & 6.80 & -- &--&--&--&--&--&--&--&--& 3 & 6 \\
NGC4395 & [Fe II] & 1.864 & 7.90 & -- &--&--&--&--&--&--&--&--& 3 & 2 \\
NGC4395 & [Cr II] & 1.869 & 6.80 & -- &--& 0.012 &--&--&--&--&--&--& 3 & 2 \\
NGC4395 & H & 1.876 & 0.00 & paschen-$\alpha$(4-3) & 1.254 &--& 0.012 & -22390 & 0 & 160 & 0 & 629.9 & 3 & 1 \\
NGC4395 & H2 & 1.892 & 0.00 & (1-0)S(4) & 0.005 &--& 0.001 & -50 & 10 & 0 & 30 & 6.4 & 3 & 0 \\
NGC4395 & [Ni II] & 1.939 & 7.64 & -- &--&--&--&--&--&--&--&--& 3 & 4 \\
NGC4395 & H & 1.945 & 0.00 & 8-4 & 0.058 &--& 0.001 & 30 & 10 & 150 & 10 & 62.1 & 3 & 0 \\
NGC4395 & H2 & 1.958 & 0.00 & (1-0)S(3) & 0.036 &--& 0.001 & -10 & 0 & 90 & 0 & 27.9 & 3 & 0 \\
NGC4395 & [Si VI] & 1.964 & 166.80 & -- & 0.056 &--& 0.002 & -100 & 0 & 140 & 0 & 24.7 & 3 & 0 \\
NGC4395 & [Fe II] & 1.967 & 7.90 & -- &--& 0.102 & 0.034 &--&--&--&--& 0.1 & 3 & 0 \\
NGC4395 & H2 & 2.034 & 0.00 & (1-0)S(2) & 0.014 &--& 0.001 & -10 & 10 & 80 & 10 & 21.2 & 3 & 0 \\
NGC4395 & H2 & 2.122 & 0.00 & (1-0)S(1) & 0.042 &--& 0.000 & -10 & 10 & 60 & 10 & 102.2 & 3 & 0 \\
NGC4395 & H & 2.166 & 0.00 & brackett-$\gamma$(7-4) & 0.087 &--& 0.001 & -30 & 10 & 110 & 10 & 104.4 & 3 & 0 \\
NGC4395 & H2 & 2.223 & 0.00 & (1-0)S(0) & 0.017 &--& 0.001 & 0 & 10 & 100 & 10 & 21.6 & 3 & 0 \\
NGC4395 & H2 & 2.248 & 0.00 & (2-1)S(1) & 0.004 &--& 0.000 & -20 & 10 & 20 & 20 & 10.7 & 3 & 0 \\
... & ... & ... & ... & ... & ... & ... & ... & ... & ... & ... & ... & ... & ... \\
\enddata

\tablecomments{The complete table is presented in the online version of the {\it Astrophysical Journal}. Here we present the first few rows to show its form and content. The measured quantities provided here are derived from the multi-component Gaussian fits described in Section~\ref{sec:methods}. We define the line as detected if the integrated flux of a best-fit single-Gaussian model has a ${\rm S/N}\geq 3$; upper limits are provided for undetected emission lines. The ``Warning'' column identifies issues with the spectra (blended feature, bad pixel, etc). 0 $-$ good fit; measurements reported. 1 $-$ primary line from deblended feature; measurements reported. 2 $-$ constrained line from deblended feature; measurements reported. 3 $-$ three or more features blended; no measurements to report. 4 $-$ weak or absent line near another strong line; no measurements to report. 5 $-$ bad fit due to failed fits or overly broad fit; no measurements to report. 6 $-$ bad fit or bad spectra due to ringing or other spectral features; no measurements to report. 7 $-$ bad spectra due to channel or sub-channel edge, detector gap, or noisy spectra; no measurements to report. 8 $-$ single gaussian signal-to-noise greater than 3 but peak flux of gaussian model is less than 3 sigma of the continuum noise; measurements reported. \\}
\tablenotetext{a}{Rest wavelengths from \href{https://physics.nist.gov/PhysRefData/ASD/lines_form.html}{NIST.}}
\tablenotetext{b}{Ionization potential energy from \href{https://physics.nist.gov/PhysRefData/ASD/ionEnergy.html}{NIST.}}
\end{deluxetable*}
\end{rotatetable}
\clearpage

\bibliography{reveallagn2}{}
\bibliographystyle{aasjournal}

\end{document}